\documentclass[11pt,leqno,fleqn,table]{article}

\usepackage[numbers]{natbib} 

\usepackage{tikz}
\usepackage{amssymb}
\usepackage{amsmath}
\usepackage{url}
\usepackage{graphicx}
\usepackage[backref]{hyperref} 
\usepackage{xcolor}
\usepackage{bm}       

\usepackage{enumitem} 

\usepackage{afterpage}  
\usepackage{float} 
\floatstyle{boxed} 

\usepackage[scaled=0.9]{helvet} 

\newtheorem{theorem}{Theorem}

\newtheorem{restxxx}[theorem]{Restriction}

\newtheorem{agreexxx}[theorem]{Agreement}

\newtheorem{termxxx}[theorem]{Terminology}

\newtheorem{notxxx}[theorem]{Notation}

\newtheorem{assumxxx}[theorem]{Assumption}

\newtheorem{convenxxx}[theorem]{Convention}

\newtheorem{exaxxx}[theorem]{Example}

\newtheorem{exexxx}[theorem]{Exercise}

\newtheorem{remxxx}[theorem]{Remark}

\newtheorem{openxxx}[theorem]{Open Problem}
\newtheorem{conjxxx}[theorem]{Conjecture}

\newtheorem{defxxx}[theorem]{Definition}
{\hfill\QED\end{defxxx}}
\newtheorem{procxxx}[theorem]{Procedure}
{\hfill\QED\end{procxxx}}

\newcommand{\HItxt}[1]{{\colorbox{blue!20}{#1}}}

\newenvironment{custommargins}[2]%
  {\addtolength{\leftskip}{#1}\addtolength{\rightskip}{#2}}{\par}

\newtheorem{Prxxx}[theorem]{Proof}
{\end{Prxxx}} 

\newcommand{\semantics}[1]{[\kern-.42em[\;#1\;]\kern-.42em]}

\newcommand{\Let}[3]%
    {\textbf{\textsf{let}}\ {#1} {#2}\ \textbf{\textsf{in}}\;{#3}\;}
\newcommand{\LET}[3]%
    {\textbf{\textsf{let}}^*\ {#1} {#2}\ \textbf{\textsf{in}}\;{#3}\;}

\newcommand{\ie}{\textit{i.e.}}
\newcommand{\eg}{\textit{e.g.}}
\newcommand\QED{$\square$} 

\newcommand{\Hide}[1]{}

\newif
\ifnote
\notetrue
\newcommand{\note}[1]
    {\ifnote {\color{red} [ #1 ] } \else {} \fi}

\begin{document}

\title{Mathematical Logic in Computer Science}

\date{June 1, 2017\qquad (last update: February 1, 2018)}

\author{Assaf Kfoury}

\maketitle

\newcommand\yrTL[5]{%
\parbox[b]{1.3cm}{\hfill{\color{blue}\bfseries\sffamily #1~$-$~}}%
\makebox[0pt][c]{\color{blue}$\bullet$}\vrule\quad%
\parbox[c]{3.6cm}{\vspace{7pt}\raggedright #2 \ \\[7pt]}%
\quad\parbox[c]{4.2cm}{\vspace{7pt}\raggedright #3 \ \\[7pt]}%
\quad\parbox[c]{4.2cm}{\vspace{7pt}\raggedright #4 \ \\[7pt]}%
\quad\parbox[c]{4cm}{\vspace{7pt}\raggedright #5 \ \\[7pt]}%
\ \\[-.2ex]
}
\newcommand\yrTLL[4]{%
\parbox[b]{1.3cm}{\hfill{\color{blue}\bfseries\sffamily #1~$-$~}}%
\makebox[0pt][c]{\color{blue}$\bullet$}\vrule\quad%
\quad\parbox[c]{4.9cm}{\vspace{7pt}\raggedright #2 \ \\[7pt]}%
\quad\parbox[c]{4.9cm}{\vspace{7pt}\raggedright #3 \ \\[7pt]}%
\quad\parbox[c]{4.9cm}{\vspace{7pt}\raggedright #4 \ \\[7pt]}%
\ \\[-.2ex]
}
\newcommand\yrTLLL[3]{%
\parbox[b]{1.3cm}{\hfill{\color{blue}\bfseries\sffamily #1~$-$~}}%
\makebox[0pt][c]{\color{blue}$\bullet$}\vrule\quad%
\quad\parbox[c]{4.1cm}{\vspace{7pt}\raggedright #2 \ \\[7pt]}%
\quad\parbox[c]{10.9cm}{\vspace{7pt}\raggedright #3 \ \\[7pt]}%
\ \\[-.2ex]
}
\newcommand\TL{%
\parbox[b]{1.30cm}{\hfill{\color{blue}\bfseries\sffamily}}%
\makebox[.4pt][c]{$\Big\vert$}\quad%
\parbox[c]{3.9cm}{\vspace{7pt}\raggedright \ \\[7pt]}%
\quad\parbox[c]{3.9cm}{\vspace{7pt}\raggedright \ \\[7pt]}%
\quad\parbox[c]{3.9cm}{\vspace{7pt}\raggedright \ \\[7pt]}%
\quad\parbox[c]{3.9cm}{\vspace{7pt}\raggedright \ \\[7pt]}%
\ \\[-.2ex]
}
\newcommand\Headings[4]{%
\hspace{1.6cm}
\parbox[c]{3.6cm}{\vspace{7pt}\color{blue}\raggedright\sffamily #1 \ \\[7pt]}%
\quad\parbox[c]{4.2cm}{\vspace{7pt}\color{blue}\raggedright\sffamily #2 \ \\[7pt]}%
\quad\parbox[c]{4.2cm}{\vspace{7pt}\color{blue}\raggedright\sffamily #3 \ \\[7pt]}%
\quad\parbox[c]{4cm}{\vspace{7pt}\color{blue}\raggedright\sffamily #4 \ \\[7pt]}%
\ \\
}
\newcommand\Headingss[3]{%
\hspace{1.9cm}
\parbox[c]{4.8cm}{\vspace{7pt}\color{blue}\raggedright\sffamily #1 \ \\[7pt]}%
\quad\parbox[c]{5cm}{\vspace{7pt}\color{blue}\raggedright\sffamily #2 \ \\[7pt]}%
\quad\parbox[c]{5cm}{\vspace{7pt}\color{blue}\raggedright\sffamily #3 \ \\[7pt]}%
\ \\
}
\newcommand\Headingsss[2]{%
\hspace{1.5cm}
\quad\parbox[c]{4.2cm}{\vspace{7pt}\color{blue}\raggedright\sffamily #1 \ \\[7pt]}%
\quad\parbox[c]{5.2cm}{\vspace{7pt}\color{blue}\raggedright\sffamily #2 \ \\[7pt]}%
\ \\
}

\section{Introduction}
\label{sect:introduction}

Others have written about the influences of mathematical logic on
computer science.  I single out two articles, which I have read and
re-read over the years:
\begin{enumerate}[itemsep=2pt,parsep=2pt,topsep=2pt,partopsep=0pt] 
\item ``Influences of Mathematical Logic on Computer Science,'' 
       by M. Davis~\cite{davis88},
\item  ``On the Unusual Effectiveness of Logic in Computer Science,''
       by J. Halpern, R. Harper, N. Immerman, P. Kolaitis, M. Vardi, and V. 
       Vianu~\cite{kolaitis-et-al2001}.
\end{enumerate}
The first of these two articles takes stock of what had already become
a productive cross-fertilization 
by the mid-1980's; it is one of several interesting articles of historical
character by M. Davis~\cite{davis1982,davis1987,davis2001} which all
bring to light particular aspects of the relationship between the two fields. 
The second article gives an account of this relationship in
five areas of computer science by the year 2000.  More on the second
article, denoted by the acronym UEL, in Section~\ref{sect:kolaitis-et-al} below.

I wanted to write an addendum to the two forementioned articles, in
the form of a timeline of significant moments in the history relating
the two fields, from the very beginning of computer science in the
mid-1950's till the present. The result is this paper. One way of
judging what I produced is to first read the penultimate section
entitled `Timeline', Section~\ref{sect:timeline} below, and then go
back to earlier sections whenever in need of a justification for one
of my inclusions or one of my omissions.

\textbf{Disclaimer:} 
This is not a comprehensive history, not even an attempt at one, of 
mathematical logic in computer science. This is a personal account of
how I have experienced the relationship between the two fields since
my days in graduate school in the early 1970's. 
So it is a personal perspective and I expect disagreements.

\textbf{Notation and Organization:}
I use italics for naming areas and topics in mathematical logic and
computer science, for book titles, and for website names; I do not use italics for
emphasis. Single quotes are exclusively for emphasis, and double quotes
are for verbatim quotations. I pushed all references and, as much as possible,
all historical justifications into footnotes.

\textbf{Acknowledgments:}
  I updated the text whenever I received comments from colleagues who
  took time to read earlier drafts.  Roger Hindley, Aki Kanamori, and
  Pawel Urzyczyn provided documents of which I was not
  aware. I corrected several wrong dates and wrong attributions, and
  made several adjustments, some minor and some significant, after
  communicating with Martin Davis, Peter Gacs, Michael Harris,
  Roger Hindley, Aki Kanamori, Phokion Kolaitis,
  Leonid Levin, Pawel Urzyczyn, and Moshe Vardi. I owe special thanks
  to all of them.

     \Hide{
\ \\ \ \medskip
     {\note{Short paragraph still missing here.}}

     Micahel Harris, Roger Hindley, Aki Kanamori, and
     Pawel Urzyczyn provided me with documents
     }
     
\Hide{ Martin Davis, Peter Gacs, Michael Harris,
  Roger Hindley, Aki Kanamori, Phokion Kolaitis,
  Leonid Levin, and Pawel Urzyczyn.

}

\Hide{ I also ignored a few suggestions for the main text, though still
  noted in footnotes, not so much out of disagreement, but out of a need to
  keep the article from mushrooming to a size I could not manage well.

}



\section{Areas of Mathematical Logic and How Far They Extend}
\label{sect:areas}

Mathematical logic is often divided into four major areas:%
  \footnote{\label{foot:four-areas}
  These are, for example, the four areas identified in a  
  handbook~\cite{barwise1977},
  and again in a very nice article~\cite{buss-et-al2001}. 
  In the latter article's first two sections, there is a discussion of 
  the interaction between mathematical logic and
  computer science. The authors are four eminent mathematical logicians.
  It is interesting that they give the lion's share
  of this interaction to \emph{computability} and \emph{proof theory},
  leaving relatively little to \emph{model theory}, and almost nothing
  explicitly to \emph{set theory}.  }
\begin{itemize}[itemsep=2pt,parsep=2pt,topsep=2pt,partopsep=0pt] 
\item \emph{computability} or \emph{recursion theory},
\item \emph{model theory},
\item \emph{proof theory},
\item \emph{set theory}.
\end{itemize}
Of these four, \emph{computability} has had the strongest
impact on the younger discipline of computer science. Arguably,
much of \emph{computability} has been taken over by researchers in
departments of computer science (or `informatics' in Europe), 
though with a difference of
emphasis. Computer scientists usually focus on \emph{tractability} or
\emph{feasible computability}, while mathematical logicians usually
focus on computability as a more theoretical concept with concerns
such as non-computability and {degrees of unsolvability}.  The
term \emph{theory of computation} usually refers to more practical
sub-areas of \emph{computability} -- some even outside 
mathematical logic proper, and closer to \emph{combinatorics}, 
\emph{number theory}, and \emph{probability theory} -- which computer
scientists have made their own and almost exclusively developed in
recent decades, \eg, \emph{randomness in computation},
\emph{resource-bounded computation}, \emph{combinatorial complexity}, 
the \emph{polynomial hierarchy},
and many refinements of the preceding with an eye on real-world applications.

Of the four areas of mathematical logic, it is fair to say that
\emph{set theory} has contributed the least to computer science, or 
that a good deal of it has played no role in computer science (\eg, think
of the very significant part of \emph{set theory} that deals with 
\emph{large cardinals}).%
   \footnote{For the purpose of a classification, I place \emph{type
   theory} under \emph{proof theory}, not \emph{set theory}.  Although
   the roots of \emph{type theory} lie in \emph{set theory}, it is
   today, at least in computer science, the study of logical systems
   that have at their core Alonzo Church's lambda calculus, which play
   a central role in the foundations of programming languages,
   computational logic, and automated proof-assistants. J.L. Bell
   traces the history of types and type theory from their beginnings
   in \emph{set theory}, around the turn of the 20th Century, to their
   gradual migration to other parts of mathematical
   logic~\cite{bell2012}.}
Nonetheless, there are parts
of \emph{set theory} as usually understood 
that have provided computer science with the means to reason about
infinite sets and infinite sequences precisely (\eg, the notions
of \emph{well-ordering}, \emph{quasi-well-ordering},
\emph{well-foundedness}, \emph{non-well-foundedness}, and similar notions,
or the principles of \emph{simple induction},
\emph{transfinite induction}, \emph{coinduction}, and various
\emph{fixed-point} notions),%
  \footnote{These notions have been studied in details by D. Sangiorgi.
  See, in particular, his fascinating historical
  account~\cite{sangiorgi2009}, which relates the notions across
  different areas of mathematical logic and computer science.
  Some of this discussion is reproduced in Sangiorgi's 
  textbook~\cite{sangiorgi2012} and in the collection of
  articles~\cite{sangiorgi+rutten2012} which he co-edited with J. Rutten.}
although these are primarily formal means for a rigorous discipline rather 
than results with potential applications.

It is also fair to say that \emph{model theory} and \emph{proof
theory} have played more prominent roles in computer science
than \emph{set theory}, and
increasingly so.  Although in the early decades of computer science,
their influence was limited and mostly theoretical, much like that of
\emph{set theory}, this is no longer the case.  Over time, their
importance increased considerably, mostly from the vantage point of
real-world applications. Early on, \emph{finite model theory},
particularly in its subarea \emph{descriptive complexity}, became the
purview of theoretical computer scientists, much less that of
mathematicians, though it called for a~theoretician's kind of
expertise and interest.%
  \footnote{Two textbooks I am familar with, by two prominent
  researchers in this area: N. Immerman~\cite{immerman1999}
  and L. Libkin~\cite{libkin2004}. An earlier comprehensive coverage
  is in a textbook by H.-D. Ebbinghaus and J. Flum~\cite{ebbinghaus+flum1995}.}
  If database management systems now form the backbone of `big data'
  applications, then \emph{finite model theory}, by way of its
  intimate synergistic relationship with \emph{database theory},
  acquires a practical dimension well beyond its intrinsic theoretical
  importance.%
  \footnote{
  A collection of articles that span \emph{finite model theory},
  from its theoretical foundations to its applications, 
  is~\cite{gradel2007}.  
  }
 
Likewise, the early impact of \emph{proof theory} on computer science
was mostly theoretical, \eg, the \emph{lambda calculus}, \emph{type
theory}, and \emph{substructural logics} came to play a central role
in the foundations of programming languages.%
  \footnote{I include several topics under this heading, although
  not always presented in a proof-theoretic framework, but more often
  in the context of the semantics of programming languages and,
  more specifically, \emph{functional} programming languages.
  Some of these topics are in the textbooks by B.C. Pierce~\cite{pierce2002},
  C.A. Gunter~\cite{gunter1992}, and J.C. Mitchell~\cite{mitchell1996},
  in a collection of articles
  edited by C.A. Gunter and J.C. Mitchell~\cite{gunter+mitchell1994},
  and in another collection edited by B.C. Pierce~\cite{pierce2005}.}
Later, they provided the foundations for most of the successful
general-purpose automated proof-assistants.%
  \footnote{A comprehensive account of these proof systems based on
  \emph{typed lambda calculi} is in a recent textbook by 
   R. Nederpelt and H. Geuvers~\cite{nederpelt+geuvers2014}.}
And later also, various \emph{deductive systems} became the
essential background for the development of a wide variety of
automated and semi-automated tools for formal verification and
satisfiability, with enormous practical consequences.%
  \footnote{Two recent book accounts of methods used in SAT and SMT solvers are:
  one by D. Kroening and O. Strichman~\cite{kroening+strichman2008}
  and one by U. Sch\"oning and J. Tor\'an~\cite{schoning+toran2013}.
  Implementation details of two highly successful SAT/SMT solvers, Z3 and CVC4,
  can be collected from their respective websites.
}

It is a pointless exercise to try to demarcate precisely the
boundaries of these areas within mathematical logic, or the boundaries
between any of these areas and other parts of mathematics, if only
because of the many overlaps.  Nonetheless, even though there are no sharp
boundaries, we do not hesitate to say, after reading an article,
that it is ``more about model theory than about proof theory,'' or
``more about logic than about topology,'' or the other way around,
etc., and so the distinctions are indeed meaningful.

For the ambiguities of where to draw these boundaries, consider the
case of \emph{category theory} and how it relates to computer science.
Initial concepts of \emph{category theory} were invented in the 1940's
and 1950's by mathematicians outside logic (homological algebra and
closely related areas in topology). Many years later,
\emph{category theory} found its way into logic, and now there is
an area called \emph{categorical logic} with many applications
in computer science.%
    \footnote{Some even think the future of 
    \emph{categorical logic} cannot be dissociated from computer science: 
    ``The fate of categorical logic is presently intimately tied
      to theoretical computer science,'' from the last paragraph in 
     a survey by two prominent categorical 
     logicians~\cite{marquis+reyes2012}. }
So, where does \emph{categorical logic} 
fit in the traditional four-area division of mathematical logic?

If we stick to the four-area division in the opening paragraph, we may
have to place (or perhaps `shoehorn') \emph{categorical logic}
somewhere in \emph{proof theory} or \emph{set theory} because of its 
extensive use of types and topos.%
    \footnote{Consider, for example, the table of contents in Bart
    Jacobs' book, \emph{Categorical Logic and Type
    Theory}~\cite{jacobs1999}. The topics in the last six, and more
    advanced, chapters of the book include: effective topos, internal
    categories, polymorphic type theory, advanced fibred categories,
    first-order dependent type theory, higher-order dependent type
    theory -- which give a sense of how deeply ingrained
    types, topos, and related notions, are in \emph{categorical logic}. }
But there are also other aspects that make \emph{categorical logic}
intersect with \emph{model theory} in a larger sense (\eg, the model
theory of typed lambda calculi takes us into the study of \emph{cartesian
closed categories}). In fact, \emph{categorical logic} cuts across 
\emph{proof theory}, \emph{set theory}, and \emph{model theory} -- 
and even \emph{computability theory}.%
   \footnote{In the four-part
    \emph{Handbook of Mathematical Logic}~\cite{barwise1977},
    there are two survey articles related to \emph{categorical logic}:
    ``Doctrines in Categorical Logic'' by A. Kock and 
    G.E. Reyes~\cite{kock+reyes1977}, which is placed  in 
    \textsc{Part A: Model Theory}, and ``The Logic of Topoi''
    by M. Fourman~\cite{fourman1977}, which is placed in 
    \textsc{Part D: Proof Theory}. In the textbook 
    \emph{Category Theory for Computing Science} 
    by M. Barr and C. Wells~\cite{barr+wells1998}, several sections in 
    different chapters include a categorical 
    treatment of functional programs and computable functions.
} 
With its recognizably distinct concepts and conventions,
it would be easier to identify \emph{categorical logic} 
as another area of mathematical logic, separate from the four 
major ones, and to consider its impact on computer science
separately.%
  \footnote{There are
  many deep interactions between the four traditional areas, so that
  a presentation of one cannot avoid broaching elements from the 
  other three, while \emph{categorical logic} stands apart in that it
  can be omitted altogether from the presentation of any of the four.
  So it makes sense, for example, that
  many standard textbooks on mathematical logic include and mix material from 
  all four major areas, and then leave \emph{categorical logic} out
  entirely. The latter is typically treated in separate and more advanced 
  books. That \emph{categorical logic} stands apart is not an original 
  observation; it is noted, critically, by categorical logicians themselves
  (\eg, see the two last pages in the last section of~\cite{marquis+reyes2012}).}

  For another example of ambiguities, now the result of shifting
  boundaries over time, consider \emph{automata theory}.  Its early
  pioneers in the 1950's and 1960's were all mathematical
  logicians.%
  \footnote{Just to mention a few of the most prominent:
  Michael Rabin, Dana C. Scott, J. Richard B\"{u}chi, Calvin C. Elgot,
  Robert McNaughton, 
  and Boris A. Trakhtenbrot, who all went on to become great contributors
  to \emph{computer science} in later years.}
  While it barely registered a mention among logicians outside
  \emph{modal logics} at the time,%
  \footnote{In the four-volume
  \emph{Handbook of Mathematical Logic}~\cite{barwise1977},
  there are no chapters devoted to
  \emph{modal logics} or to \emph{automata theory} or to what relates
  the two. One chapter in the handbook, entitled ``Decidable Theories''
  by M. O. Rabin, includes the author's theorem without its proof
  on the decidability of the second-order theory
  of two successor functions, though it does also
  mention that the proof involves an ``extension of the theory of automata
  to cover the case of a finite automaton operating on an infinite tree''
  but without further explanation. In a later section in the same chapter, there
  is a passing mention that the forementioned theorem can be used to obtain
  positive decidability results for ``non-classical logics'' (here meaning
  \emph{modal logics}). In the article ``Prospects for Mathematical Logic in
  the Twenty-First Century''~\cite{buss-et-al2001}, also cited in
  footnote~\ref{foot:four-areas}, there is a single mention of `automata theory'
  in a list of over $50$ possible connections between logic and computer science
  (Table 2 on p. 179) and a single mention of `modal logics' earlier in the text,
  with no presentation of results in relation to both topics or to the deep
  connections between the two.}
  \emph{automata theory} quickly became a core part of computer
  science. It has remained so, though it has also gone through ebbs
  and flows of relevance (and, I may add, popularity) in different
  parts of the field. In the early years, it was squarely placed
  in \emph{theory of computation}; in later years, through its
  counterpart \emph{formal-language theory}, it became part of the
  essential background in the study of programming languages and
  compiler design (mostly involving the finite automata associated
  with the lower levels of the Chomsky hierarchy); in later years
  still, it acquired a special importance in the study of \emph{modal
  logics} and \emph{temporal logics} which deeply
  influenced \emph{model checking} (involving various kinds of
  automata on infinite objects); and now, there are a few who want
  to push \emph{automata theory} completely out of \emph{theory of computation}.%
  \footnote{This history is partly reflected in textbooks with titles
  containing `theory of computation' or `introduction to computation',
  or which are presented in their introductory chapters as covering such
  material.  At least a third of the chapters in several standard
  texts~\cite{hopcroft+ullman+motwani2006,
  lewis+papadimitriou1997,sipser2012} written up until the late 1990's
  (and beyond in the newer editions) are, partially or entirely,
  on \emph{automata theory}. In more recent years, \emph{automata
  theory} has fallen into disfavor among some people;
  the disfavor extends to at least the part on finite
  automata, not quite to automata on infinite inputs, although the
  former is arguably a prerequisite for the latter.  Consider the
  following comments in a recent book by a prominent theoretical computer
  scientist~\cite[page 36]{goldreich2008}:
  \begin{quote}
  ``We reject the common coupling of computability theory with the theory
    of automata and formal languages. Although the historical links between
    these two theories (at least in the West) cannot be denied, this fact
    cannot justify coupling two fundamentally different theories (especially
    when such a coupling promotes a wrong perspective on computability
    theory). Thus, in our opinion, the study of any of the lower levels of
    Chomsky's Hierarchy~\cite[Chap. 9]{hopcroft+ullman+motwani2006}
    should be decoupled from the study of
    computability theory (let alone the study of Complexity Theory).''
  \end{quote}
  Is there more than meets the eye in such a categorical opinion? The
  book leaves out \emph{automata theory} in all its aspects. Ignored
  by such an opinion is any recognition that the notion of
  \emph{nondeterminism}, though fundamental in the definition of
  complexity classes studied in this book, was historically introduced
  in \emph{automata theory} (first in~\cite{rabin+scott1959}).
  }
  However, if only by way of its importance to the development
  of \emph{model checking} and the latter's great successes in practice
  (\eg, the verification, usually automated, of software
  properties against their formal specification), \emph{automata theory}
  should occupy pride of place among works of mathematical logicians
  that have had deep repercussions in computer science.%
  \footnote{A textbook survey of methods of \emph{model checking} in practical
  applications is by C. Baier and J.-P. Katoen~\cite{baier+katoen2007}. 
  A~book that deals with implementation issues of a particular model-checker
  (Spin) is by M. Ben-Ari~\cite{ben-ari2008}. A collection of
  papers exploring many different aspects relating
  mathematical logic and \emph{automata theory} is~\cite{flum+gradel+wilke2008},
  and another collection of papers on \emph{modal logics}
  is~\cite{blackburn+benthem+wolter2006}, which includes a survey by M.Y. Vardi
  of results  connecting automata (on infinite inputs) and
  modal logics~\cite{vardi2006}.  
  }


\section{`On the Unusual Effectiveness of Logic in Computer Science' }
\label{sect:kolaitis-et-al}

The paper whose title is the title of this section gives an account of the
relationship between the two fields as it stood around the year 
2000.%
  \footnote{The title of this paper~\cite{kolaitis-et-al2001} is
  probably inspired by earlier articles on the ``unusual'' or
  ``unreasonable effectiveness'' of mathematics in the natural
  sciences.  Among these, there are two (and perhaps others), one by
  E. Wigner~\cite{wigner1960} and one by
  R.W. Hamming~\cite{hamming80}, whose examination is easily
  redirected to be about the importance of mathematics and mathematical
  logic in computer science.}%

That paper (which I denote by the acronym UEL), authored by six
theoretical computer scientists, surveys five areas of
computer science where mathematical logic figures most
prominently. The titles of the relevant sections are (here numbered
according to their order in UEL):
\begin{enumerate}[itemsep=2pt,parsep=2pt,topsep=2pt,partopsep=0pt] 
\item[2.] Descriptive Complexity 
\item[3.] Logic as a Database Query Language 
\item[4.] Type Theory in Programming Language Research 
\item[5.] Reasoning About Knowledge 
\item[6.] Automated Verification of Semiconductor Designs 
\end{enumerate}
All of these five sections use a profusion of elements from \emph{model theory} 
and \emph{proof theory}, in different degrees to be sure; to a much lesser extent,
elements from \emph{computability theory}; rather little, at least explicitly,
from \emph{categorical logic}; and nothing from the deeper parts
of \emph{set theory}. Section 2 in UEL can be placed
under \emph{finite model theory} and is also perhaps the one that uses
the most notions from \emph{computability theory}; Section 3 is mostly
about first-order definability in finite relational structures and can
be therefore placed under \emph{finite model theory}; Section 4 can be viewed
as an explanation for the power of the Curry-Howard Isomorphism (also
known as the \emph{propositions-as-types} principle) in functional
programming and is therefore closely related to \emph{proof theory},
and can be read as the closest to \emph{categorical logic}; Sections 5
and 6 in UEL use modal logics and their semantics (Kripke structures in
Section 5, Linear Temporal Logic in Section 6) and can be construed as
involving primarily elements of both \emph{model theory} and
\emph{proof theory}.

At the very end of UEL, the authors write:
\begin{quote}
  ``The effectiveness of logic in computer science is not by any
  means limited to the areas mentioned here. As a matter of fact, it
  spans a wide spectrum of areas, from artificial intelligence to
  software engineering. Overall, logic provides computer science with
  both a unifying foundational framework and a powerful tool for
  modeling and reasoning about aspects of computation.''
\end{quote}
That conclusion is as much in force today as it was two decades ago.
However, if the aim was to select areas of computer science
where mathematical logic had demonstrated its strongest impact, 
then there was at least one conspicuous omission in UEL: 
the extensive body of research in the area \emph{logics of programs}
which, in both quantity and depth, had an equal or stronger
claim for showcasing the close relationship between the two fields
by the year 2000.%
  \footnote{A survey of \emph{logics of programs} up to the late 1980's
  is by D. Kozen and J. Tiuryn~\cite{kozen+tiuryn1989}, and a book-length account 
  of \emph{dynamic logic} (which is a part of \emph{logics of programs})
  up to the late 1990's is by D. Harel, D. Kozen, 
  and J. Tiuryn~\cite{harel-et-al}.
  }

Of course, unbeknownst to the authors of UEL were the
unprecedented advances yet to be experienced after the year 2000.
Though the results in UEL were interesting in their own
right (for someone versed in mathematical logic) and significant in
their respective areas, it is also fair to say their effect
outside -- say, in systems areas (\eg, operating systems) or in
application areas (\eg, machine learning) -- had been rather
limited or nil. After UEL's publication, changes in the relationship between 
mathematical logic and computer science, not just incremental but some 
truly transformative, have taken place in rapid succession -- as I try to 
relate below.


\section{Scope of an Ever Deeper Integration}
\label{sect:integration}

I make a selection of events that illustrate 
the impact of mathematical logic on the younger discipline over a longer time
span, from the mid-1950's until the present. I want to stress my
initial disclaimer: My aim is to record significant turning points
and moments of recognition, as I see them, not to produce an
exhaustive chronology.

I divide the development of computer science into three 
periods, each of about 20 years. This division is rather arbitrary, 
but it makes my presentation a little easier: the formative years (1953-1975), 
the consolidation years (1976-1995), and the mature or growth years (1996-2017).%
   \footnote{\label{foot:beginnings}
   And what I mark as the beginnings of computer science in
   the 1950's is not recognized by everyone. Others like to say the
   birth of computer science was some two decades earlier, in the 1930's:
   ``When Turing came to Princeton to work with Church, in the orbit of
   G\"{o}del, Kleene, and von Neumann, among them they founded a field of
   computer science that is firmly rooted in
   logic''~\cite{appel2012}. Put that way, mathematical logicians are
   made the true and sole progenitors of computer science -- an
   assertion which, I suspect, will strike many (most?)
   computer scientists as a bit of a hyperbole.
   More emphatically in a similar vein, a prominent
   \emph{theory-of-computation} researcher marks 1936 as the beginning:
   ``In this extremely readable paper~\cite{turing1936}, Turing
    gave birth to the discipline of Computer Science, ignited the computer
    revolution which radically transformed society, and $\ldots$''
    (Chapter 2 in~\cite{wigderson2017}).}

I omit connections that are strictly related to \emph{computability}, 
including those under the more restrictive rubric \emph{theory of computation},
the sub-areas dominated by theoretical computer scientists.  At least one 
good reason for this omission: There are just too many 
\emph{computability}-related milestones which have permeated computer 
science from the very beginning, perhaps to the point of crowding out 
other important contributions of mathematical logic in the mind of many.%
   \footnote{And others are more qualified than I to write a survey 
   of \emph{computability}-related milestones in computer science.}

My focus is therefore on connections between computer science and parts
of mathematical logic that are commonly considered in (using the headings
of the four-part division in Section~\ref{sect:areas}): 
\begin{quote}
  \emph{model theory}, 
  \emph{proof theory}, and \emph{categorical logic} -- \\ leaving out
  parts that are primarily in \emph{computability theory}.%
     \footnote{\label{trackA-trackB} The term \emph{formal methods}, used by many
     computer scientists, roughly corresponds to my focus in this paper, 
     which is also roughly the focus of Track B in the journal
     \href{https://www.journals.elsevier.com/theoretical-computer-science/}{\em
     Theoretical Computer Science} (TCS) and ICALP conferences of
     the \href{http://eatcs.org/index.php/conferences}
     {\it European Assoc. for TCS} (EATCS).
     Quoting from the latter website, ``Track A [$\ldots$] correspond 
     to Algorithms, Automata, Complexity, and Games, 
     while Track B to Logic, Semantics, and Theory of Programming.'' 
     Following this usage, an alternative title for this paper could be 
     `A Perspective on Formal Methods in Computer Science' or 
     `A Perspective on Logic, Semantics, and 
     Theory of Programming in Computer Science', instead of
     `Mathematical Logic in Computer Science'. At the end I chose
     the latter title to avoid some of the limitations suggested by the
     former. 
     }
\end{quote}
However, there are famous results about limits of \emph{computability} that
should be mentioned here (\eg, \emph{NP-completeness}) because they
shed light on limits (as well as successes in bypassing these limits)
in applications of \emph{model theory} and \emph{proof theory} decades
later (\eg, SAT/SMT solvers and model-checkers).%
   \footnote{I should add that my focus is in harmony with  
    UEL's focus~\cite{kolaitis-et-al2001}, as presented in
    Section~\ref{sect:kolaitis-et-al} above. 
    There is no section in UEL devoted to a topic in 
    \emph{computability} proper, although there are many such
    topics that have had major impacts on computer
    science or that have been developed by
    computer scientists before the year 2000, around the time of
    UEL's publication (\eg, the theory of \emph{alternating Turing machines}
    or the many topics related to \emph{subrecursive hierarchies}). 

    My focus is also in harmony with the selection of topics in the
    \emph{Handbook of Logic in Computer Science}~\cite{abramsky-et-al}.
    Out of 24 chapters in the first five volumes of the \emph{Handbook},
    at least 21 chapters deal primarily with issues related to 
    first-order model theory and universal algebra,
    category theory and topology, domain theory and 
    denotational semantics, types, modal logics, rewriting systems and
    process algebras -- this information can be gathered by reading 
    titles and introductions -- which are 
    all topics with considerable overlaps with
    traditional \emph{model theory}, \emph{proof theory}, and \emph{categorical
    logic}. There is one chapter on \emph{recursion theory} and one chapter on
    \emph{complexity of logical theories}. Outside the chapter on 
    \emph{recursion theory},
    there is arguably no chapter on a topic that can be placed mainly
    under \emph{computability}/\emph{recursion theory}
    or the narrower \emph{theory of computation}, and no chapter on a topic
    that is mainly under \emph{set theory}.

    And my focus is again in harmony with the selection of topics
    in textbooks with titles such as
    \emph{Logic for Computer Science}~\cite{reeves+clarke2003,
    ben-ari2012,gallier2015}
    and \emph{Logic in Computer Science}~\cite{huth+ryan2004},
    whose contents are mostly a mixture of \emph{model theory} and 
    \emph{proof theory} (and a little of \emph{categorical logic}
    when dealing with types or commutative diagrams).}


\subsection{First Two Decades} 
\label{sect:first-twenty}

Figure~\ref{fig:first-twenty} is a timeline of relevant moments during
the first twenty years (which I stretched by including an event in
1953), the formative decades of computer science as a separate
discipline.

During these early years, there is relatively little to relate from
\emph{model theory}, \emph{proof theory}, and \emph{categorical logic}
-- and mathematical logic in general outside \emph{theory of computation} --
to important milestones in computer science (highlighted
with a gray background in Figure~\ref{fig:first-twenty}). In the
column with the heading `Milestones/Accolades', I choose to highlight
four:
\begin{itemize}[itemsep=2pt,parsep=2pt,topsep=2pt,partopsep=0pt] 
\item The first implementation of Lisp (1959), a functional programming
  language influenced by Alonzo Church's lambda-calculus.
\item The invention of Hoare Logic (1969), which should be considered
  the first precisely formulated \emph{logic of programs}.%
  \footnote{Others consider work by J. McCarthy~\cite{mccarthy1963}
    and R.W. Floyd~\cite{floyd1967}, both preceding C.A.R. Hoare's
    paper~\cite{hoare1969}, as the true beginning
    of logics of programs. And some use the name
    `Floyd-Hoare Logic' instead of `Hoare Logic'. Although these two papers
    suggested, if only implicitly, the seminal ideas of \emph{loop
      invariants} and \emph{correctness assertions}, these were not
    incorporated yet in the rules of a deductive system. Moreover, the
    programming formalisms used by McCarthy (recursive definitions in
    a functional-programming style) and Floyd (flowcharts) hampered,
    arguably, the adaptation of their respective approaches to other programming
    formalisms in later years. Not every programming formalism can be
    viewed as a (straightforward) adaptation of recursive definitions,
    or can be translated into flowcharts. From the early 1970's
    to the late 1990's and beyond, there was a large body of
    research (the theory of 
    \emph{program schemes}, initiated by D. Luckham, D.M.R. Park, and
    M.S. Paterson~\cite{luckham+park+paterson1970}, and their 
    collaborators and students~\cite{paterson+hewitt1970,
    garland+luckham1973,kfoury1974,kfoury+park1975},
    which, among other problems, analyzed program formalisms depending on whether 
    programs are or are not flowchartable (not all are~\cite{kfoury+urzyczyn1985}).}
\item The programming language Pascal (1970), the first imperative
    and procedural PL to allow
    programmers to define higher-order (\ie, nested to any depth level)
    structured datatypes and procedures.%
    \footnote{Support of higher-order procedures was included in Pascal's
    language definition, but not in its implementations, which were typically
    limited to second order. Pascal was also meant to be strongly-typed,
    but wasn't quite, the most notorious
    hole in its type system being with variant records:
    ``There is a large hole in the type-checking near variant records, 
     through which some otherwise illegal type mismatches can be obtained,''
    as pointed out early on by a Unix's co-developer, Brian W. 
    Kernighan~\cite{kernighan1981}.
    } 
\item The Cook-Levin Theorem on NP-completeness (1971).
  Stephen Cook directly related his theorem to the complexity
  of automated theorem-proving (though there was no tool at the
  time comparable to a modern SAT solver), 
  while Leonid Levin formulated it
  in relation to search problems.%
  \footnote{The result is in Cook's paper~\cite{cook1971}, and published
  independently in Levin's paper~\cite{levin1973}
  (in Russian). Articles in the 1970's, and even in the 1980's and later,
  often gave credit to Cook only. Though published in 1973,
  Levin's paper had been mentioned in talks a few years before.
  A detailed history is an article
  by Boris Trakhtenbrot~\cite{trakhtenbrot1984}, which includes 
  an annotated English translation of Levin's paper.}
\end{itemize}
In the column `Major Conferences' I~highlight:
ICALP~(1972), CADE~(1974) and POPL~(1974). I~highlight
ICALP because of its Track B coverage.%
   \footnote{The history of ICALP is detailed at the
   {\href{http://eatcs.org/index.php/conferences}{\it EATCS}} website.
   See my comments on EATCS's Track B in footnote~\ref{trackA-trackB}.}
CADE started as a
workshop with a relatively small participation, becoming a full-fledged
conference with a larger attendance in the 1980's; it has the
distinction of being the first regular, annual or biennial, 
conference devoted to problems of automated formal reasoning.%
  \footnote{CADE's history can be found at its official
    website {\href{http://www.cadeinc.org/conferences}{
        \emph{Conf. on Automated Deduction}}.}}  
Even though a significant portion of POPL articles are about pragmatics and
implementation of programming languages, a good many other POPL articles
cover topics based on ideas that mathematical logicians would readily recognize 
as coming from \emph{model theory}, \emph{proof theory}, 
or \emph{categorical logic}. POPL was the first of several annual conferences 
with similar and overlapping coverages,
including PLDI (first held in 1988) and ICFP (first held in 1996).%
  \footnote{POPL's history is at two webpages:
 {\href{http://dl.acm.org/event.cfm?id=RE180}{\it ACM Digital Library: POPL}}
  and  
  {\href{https://en.wikipedia.org/wiki/Symposium_on_Principles_of_Programming_Languages}{\it Symp. on Principles of Prog. Lang}}.
  More on PLDI at: 
  {\href{https://en.wikipedia.org/wiki/Programming_Language_Design_and_Implementation}{\em Prog. Lang. Design and Implementation}}.
  More on ICFP at:
  {\href{https://en.wikipedia.org/wiki/International_Conference_on_Functional_Programming}{\em Int. Conf. on Functional Prog}}. 
 PLDI is the successor of several conferences with different names held
 in 1979 and then annually from 1982 to 1987, inclusive.
 ICFP is the successor of two conferences:
 LFP (\emph{LISP and Functional Prog.}) and
 FPCA (\emph{Functional Prog. and Comp. Architecture}); LFP was held
 from 1980 to 1994, inclusive, every two years; FPCA was first held in 1981
 and then from 1985 to 1995, inclusive, every two years.}

In the column `Major Conferences' I list but do not
highlight STOC (1969) and FOCS (1975), the twin standard-bearer
conferences in \emph{theory of computation} and, with some qualms,
outside my focus.%
  \footnote{The relation with STOC and FOCS is not clear cut. 
  It has changed over time. In the 1970's, 1980's, and at least into the 
  1990's and even later, many articles appearing in STOC and FOCS, though a minority
  of the total, were arguably more about \emph{formal methods} (\ie, EATCS's
  Track B) than about \emph{theory of computation} (\ie, EATCS's
  Track A). These were often presented in separate sessions of STOC and FOCS
  that many participants would identify as the `semantics' sessions, 
  even though their topics were not necessarily related to semantics in any 
  obvious way. That articles on formal methods  were given relatively short 
  shrift in STOC and FOCS in the early decades was doubtless an incentive 
  (not the only one, of course) for the emergence of several new workshops and 
  conferences focused on logic and formal methods,
  in the 1980's and 1990's and later.  }

I stretched the `First Two Decades' by including the Cambridge
Diploma in Computer Science (1953), in deference to its promoters'
claim that the diploma was the ``world's first''. A closer look
shows it was awarded after a one-year, master's level, course of
studies in the use of ``electronic computing-machines'' in numerical
analysis, which is rather different from the coverage of
a master's level degree in computer science (or informatics) today.%
  \footnote{The Wikipedia page 
    {\href{https://en.wikipedia.org/wiki/Cambridge_Diploma_in_Computer_Science}
      {\it Cambridge Diploma in Computer Science}} says more on what is claimed
      to be the ``world's first''. For a sense of how far computer science has
    moved from its origins in the mid-1950's (or earlier for some,
    see footnote~\ref{foot:beginnings}), at least partly because of
    the influence of mathematical logic, compare with the following:
    Every master's level student today is expected to know something
    about \emph{feasible} versus \emph{unfeasible} algorithmic
    problems (from a course in \emph{theory of computation}) or about
    \emph{types} (from a course on \emph{principles of programming
    languages}). NP-completeness was not yet known in the 1950's, so no
    comparison is possible on this, but \emph{computable functions} and
    \emph{decision problems} had been studied since at least
    the 1930's $\ldots$ . }


\subsection{Second Two Decades} 
\label{sect:second-twenty}

Figure~\ref{fig:second-twenty} is a timeline of relevant moments
during the second two decades. This is a period of consolidation,
when many departments, schools, and colleges, of computer science
 are established separately, with an identity distinct from 
engineering and other mathematical sciences. This is also a period
of greater recognition of the role of mathematical logic in computer
science, when Turing Awards are given to computer scientists working
from a distinctly formal-logic perspective:
\begin{itemize}[itemsep=2pt,parsep=2pt,topsep=2pt,partopsep=0pt] 
\item Michael O. Rabin and Dana S. Scott (1976), for their joint article
  ``Finite Automata and Their Decision Problems'' from 1959.%
   \footnote{Just as interesting as their joint article mentioned in
   the Turing Award citation~\cite{rabin+scott1959} are their separate
   Turing Award lectures:
   Rabin's lecture was about topics in \emph{theory of computation}
   (EATCS's Track A)~\cite{rabin1977}, Scott's lecture about topics 
    in \emph{semantics} and \emph{theory of programming}
    (EATCS's Track B)~\cite{scott1977}. Scott's lecture is 
    more aligned than Rabin's lecture with my focus in this paper.}
\item C.A.R. Hoare (1980), partly in recognition of his invention of
   Hoare Logic.
\item Edgar F. Codd (1981), in recognition of his contributions
   to the theory of database systems.    
\item S.A. Cook (1982), partly in recognition of his work on the
   complexity of formal proofs.
\item Robin Milner (1991), in recognition of work which is arguably
   entirely within the space created by
   \emph{model theory}, \emph{proof theory}, and \emph{categorical logic},
   as adapted to the needs of computer science.   
\end{itemize}
M.O. Rabin, D.S. Scott, and S.A. Cook, trained as mathematical
logicians and their entire careers bear witness to the deep
connections between the two fields. They studied under
two giants of mathematical logic, Alonzo Church (doctoral advisor
of both Rabin and Scott) and Hao Wang (Cook's doctoral advisor).%
  \footnote{To do full justice to Alonzo Church's contributions to
  computer science, someone else should survey not only his own 
  accomplishments but also those of his many doctoral students. 
  In addition to Rabin and Scott, they include (in no particular order): 
  Alan Turing, Stephen Kleene, Hartley Rogers,
  Martin Davis, J. Barkley Rosser (major contributor to the lambda calculus), 
  Peter Andrews (developer of the
  \href{http://gtps.math.cmu.edu/tps.html}{\em TPS automated theorem prover}), 
  John George Kemeny (designer of the
  \href{https://en.wikipedia.org/wiki/BASIC}{\em BASIC programming language}), 
  and another two dozens distinguished logicians.}
I include E.F. Codd in my list because his ``introduction of the relational
data model and of first-order logic as a database query language definitely
changed the course of history [of the database field].''
  \footnote{My quote is from a private communication with Phokion
  Kolaitis.  More on the influences of mathematical logic underlying
  Codd's work are in his citation at the
  {\href{http://amturing.acm.org/award_winners/codd_1000892.cfm} {\em
  A.M. Turing Awards}} website.}

My qualification for R. Milner above may be questioned, so I quote the
Turing Award citation in full, which says that the award resulted from
``three distinct and complete achievements,'' namely:%
  \footnote{ Robin Milner's citation is found at the
  {\href{http://amturing.acm.org/award_winners/milner_1569367.cfm}
   {\em A.M. Turing Awards}} website.}

\begin{minipage}{.94\textwidth}
\begin{enumerate}[itemsep=2pt,parsep=2pt,topsep=2pt,partopsep=0pt] 
\item[``1.] LCF, the mechanization of Scott's Logic of Computable Functions, 
      probably the first theoretically based yet practical tool for 
      machine assisted proof construction;
\item[2.] ML, the first language to include polymorphic type inference together 
      with a type-safe exception-handling mechanism;
\item[3.] CCS, a general theory of concurrency.''
\end{enumerate}
\end{minipage}

Milner's achievements $1$ and $2$ cannot be understood outside the
background of \emph{typed $\lambda$-calculi} (which 
mix \emph{rewriting}, \emph{types}, and \emph{deductive systems}, all falling
under \emph{proof theory} and \emph{categorical logic} broadly speaking),
and Milner's achievement $3$ is an effort to formalize
a Calculus of Communicating Systems in the form of a 
\emph{transition system} (again traceable back to 
\emph{proof theory} and \emph{categorical logic}) and to 
formalize the latter's semantics using
the notion of \emph{bisimulation} (akin and inspired by
back-and-forth arguments in \emph{model theory} and
\emph{set theory}).%
  \footnote{There is a vast literature by computer scientists
   on \emph{bisimulation}, \emph{coinduction}, 
   \emph{greatest fixpoint}, and related notions, starting with the 
   publication of two of D.M.R. Park's papers~\cite{park1981a,park1981b}
   in 1981 where \emph{bisimulation} is fully defined for the first time.
   My sense is that this body of research deserves a highlighted entry in my 
   timeline (if I get to produce a second, more detailed edition of the timeline)  
   for what became a highly successful and transformative framework for 
   the analysis of concurrency, infinite processes, and related notions. 
   D. Sangiorgi gives a history of these notions in a long 
   article~\cite{sangiorgi2009}, where he also discusses akin notions 
   (sometimes with different names) in \emph{modal logic} and \emph{set theory},
   and mentions in passing connections with Ehrenfeucht-Fra\"{i}ss\'{e} games
   (EF games). Forms of bisimulations can be viewed as special families of partial 
   isomorphisms, corresponding to a restricted type of EF
   game~\cite{thomas1997,flum+gradel+wilke2008}. 
   A deeper comparison between EF games 
   and \emph{bisimulation} is in several other papers, including by Martin Otto 
   and his colleagues~\cite{dawar+otto2009,gradel+otto2014}.}

The 1980's and early 1990's saw the very beginnings of three important
automated projects, which can be viewed as current standard-bearers of
proof assistants (Coq and Isabelle) and model checkers (Spin),
listed in Figure~\ref{fig:second-twenty} under the column 
`Milestones/Accolades':%
  \footnote{With apologies to colleagues who may feel that my singling
    out of Coq, Isabelle, and Spin for recognition, does not give due
    credit to their work on other automated systems in later years.}
\begin{itemize}[itemsep=2pt,parsep=2pt,topsep=2pt,partopsep=0pt] 
\item Coq (1984).%
  \footnote{Its history
    is found on the Web at
    {\href{https://coq.inria.fr/about-coq}{\emph{What Is Coq?}}}}
\item Isabelle (1986).%
   \footnote{ More on the {\href{http://isabelle.in.tum.de/}{\em Isabelle}} 
    proof-assistant from its webpage.}
\item Spin (1991).%
   \footnote{ More on Spin from the webpage
   {\href{http://spinroot.com/spin/whatispin.html}{\em Verifying
    Multi-threaded Software}}.}
\end{itemize}
The 1980's and early 1990's also mark the beginnings of several  
academic conferences devoted to various aspects of mathematical logic in
computer science, as shown under the column `Major Conferences'
in Figure~\ref{fig:second-twenty}.

Finally, I list as a major milestone the Curry-Howard Isomorphism (1980).%
\footnote{As with any concept with many threads and contributors, it is a
    little tricky to give due credit for how the Curry-Howard
    Isomorphism (CHI) and its many variations have taken shape over
    the years. For its earliest version, I quote
    from~\cite{hindley1997}, page 74: ``The CHI was first hinted at in
    print in~\cite{curry1934} (1934), and was made explicit
    in~\cite{curry1942} (1942) and in~\cite{curry+feys1958} (1958).
    But it was viewed there as no more than a curiosity.''  While
    Curry was first to notice that `types' are `theorems', it is
    probably right to say Howard in the 1960's was first to
    notice that `term reduction' is `proof normalization'.  An
    easy-to-read historical account of the CHI is by
    P. Wadler~\cite{wadler2015}, which includes an interesting email
    exchange with Howard and clarifies some of the attributions.
    Howard's paper was published in 1980~\cite{howard1980}, though it
    had been privately circulated since 1969.}
This isomorphism expresses a correspondence between two
unrelated formalisms -- \emph{proof systems} and \emph{programming formalisms} --
which asserts that the two are fundamentally the same kind of mathematical objects.
It turned out to be an extremely productive correspondence, the basis of a totally
different approach to the design of typed programming languages,
among other deep changes in both \emph{proof theory} and 
\emph{programming language theory} and in the relation between the two.%
   \footnote{Reviewing the impact of \emph{type theory}, Robert Harper
   wrote around the year 2000~\cite{kolaitis-et-al2001}:
   \begin{quote}
     ``In the 1980's and 1990's the study of programming languages was
     revolutionized by a remarkable confluence of ideas from
     mathematical and philosophical logic and theoretical computer
     science. Type theory emerged as a unifying conceptual framework
     for the design, analysis, and implementation of programming
     languages.  Type theory helps to clarify subtle concepts such as
     data abstraction, polymorphism, and inheritance. It provides a
     foundation for developing logics of program behavior that are
     essential for reasoning about programs. It suggests new
     techniques for implementing compilers that improve the efficiency
     and integrity of generated code.''
   \end{quote}
   This ``revolution'' caused by \emph{type theory}, as described by
   Harper, can be traced back and attributed to the Curry-Howard
   Isomorphism, though it did not come early enough to block the
   ravages caused by programming languages like PL/1 (1964, first
   design) -- see ``Section 8: Criticisms'' in the
   webpage \href{https://en.wikipedia.org/wiki/PL/I}{\it PL/1} . A
   thorough book-length account of the Curry-Howard Isomorphism is by
   M.H. S{\o}rensen and P. Urzyczyn~\cite{sorensen+urzyczyn2006}; a
   collection edited by Ph. de Groote~\cite{degroote1995} reproduces
   several of the seminal papers; and an interesting book, though mostly limited
   to the author's research interests in arithmetic, is by
   H. Simmons~\cite{simmons2000}.}
 

\subsection{Third Two Decades} 
\label{sect:third-twenty}

Since the mid-1990's we have witnessed truly transformative changes, 
some would say `paradigm shifts', in the relationship between the two fields.
We can now talk as much about the impact of \emph{mathematical logic} on 
\emph{computer science} as about the converse: the impact of 
\emph{computer science} on \emph{mathematical logic} (or, more broadly, on
\emph{mathematics} generally).%
   \footnote{The optimism expressed in this section about the reverse
   impact of computer science on mathematics in general is not shared
   by many mathematicians, perhaps by most outside the community of
   mathematical logicians. In particular, the idea that an interactive
   proof assistant is more than a `super calculator', and can be used
   to search for and explore alternatives, seems antithetical to
   what many profess they do when they prove a theorem. Pierre
   Deligne, a winner of the Fields Medal, says outright, ``I don't
   believe in a proof done by a computer.'' And he adds, ``I believe
   in a proof if I understand it,'' thus suggesting that the use of
   automated tools is an obstacle to understanding a proof~\cite{horgan1993}.
    }
To illustrate this converse, I single out 
five events in five different areas of mathematics
(Figure~\ref{fig:third-twenty}), triggered or made possible by
logic-based developments in computer science, with each event
deserving the distinction of being `first' in its respective area:
\begin{itemize}[itemsep=2pt,parsep=2pt,topsep=2pt,partopsep=0pt] 
\item \emph{Boolean algebra} -- a formal proof that every Robbins algebra 
      is a Boolean algebra, using the automated theorem-prover EQP (1997).%
      \footnote{The theorem asserting that \emph{every Robbins algebra is a 
      Boolean algebra} means that a set of three equations, first formulated
      by Herbert Robbins, are equivalent to the familiar equations of Boolean 
      algebra that govern unions, intersections, and complements among sets.
      A history of the problem can be found on the Web at 
      {\href{https://en.wikipedia.org/wiki/Robbins_algebra}{\it Robbins algebra}}.
      Technical details are in an article by W. McCune~\cite{mccune1997}.
      Simplifications and a particularly lucid presentation are in an article
      by B. Dahn~\cite{dahn1998}. The theorem-prover
      {\href{https://www.cs.unm.edu/~mccune/eqp/}{\it EQP}},
      used in solving the Robbins-algebra problem, was derived from the automated
      theorem-prover {\href{https://www.cs.unm.edu/~mccune/otter/}{\it Otter}}
      and developed by the same group, and the latter was more recently 
      superseded by {\href{https://www.cs.unm.edu/~mccune/mace4/}{\it Prover9}}.}
\item \emph{Graph theory} -- a formal proof of the Four-Color Theorem using 
      the automated interactive proof-assistant Coq (2008).%
      \footnote{The Four-Color Theorem asserts:
      \emph{The regions of any simple planar map can be colored with only four 
      colors, in such a way that any two adjacent regions have different colors}.
      A short presentation of the formal proof with Coq is by
      Georges Gonthier~\cite{gonthier2008}
      The original proof of the Four-Color Theorem by K. Appel and 
      W. Haken~\cite{appel+haken1977} used a computer program, but the 
      correctness of that program (not the proof method) was never completly 
      checked, namely, ``the part [in that program] that is supposedly 
      hand-checkable is extraordinarily complicated and tedious, and as far as 
      we know, no one has verified it in its entirety,'' as reported by 
      N. Robertson, D.P. Saunders, P. Seymour, and R. Thomas,
      {\href{http://people.math.gatech.edu/~thomas/FC/fourcolor.html}
            {\em The Four Color Theorem}}.
      More on Coq from its website {\href{https://coq.inria.fr/about-coq}
      {\it The Coq Proof Assistant}}. }
\item \emph{Group theory} -- a formal proof of the Odd-Order Theorem,
      also known as the Feit-Thompson Theorem, using 
      the automated interactive proof-assistant Coq (2012).%
      \footnote{The Feit-Thompson Theorem asserts: \emph{Every finite
      group of odd order is solvable.} A presentation of the formal proof 
      with Coq is by G. Gonthier \emph{et al}~\cite{gonthier2013}. 
      The {\href{https://en.wikipedia.org/wiki/Feit–Thompson_theorem}{\it
      Feit-Thompson theorem}} webpage discusses its significance for the 
      {\href{https://en.wikipedia.org/wiki/classification_of_finite_simple_groups}
       {\it Classification of finite simple groups}}. The latter is said
       to be the longest proof in the  
      {\href{https://en.wikipedia.org/wiki/List_of_long_mathematical_proofs}
      {\it List of long mathematical proofs}}.}
\item \emph{Three-dimensional geometry} -- a formal proof of the Kepler
      Conjecture on dense sphere packings using the automated proof-assistants
      HOL Light and Isabelle (2015).%
      \footnote{The Kepler Conjecture asserts: \emph{No packing of equally-sized
      spheres in Euclidean three-dimensional space has density greater than
      that of the face-centered cubic packing}. It is more than 300 years old
      and considered the oldest problem in three-dimensional geometry. A history
      of the problem is at {\href{https://en.wikipedia.org/wiki/
      Kepler_conjecture}{\it Kepler conjecture}} webpage 
      and in a paper by T. Hales \emph{et al}~\cite{hales2017}. The latter 
      paper explains more of the mathematical details than the webpage.
      }
\item \emph{Number theory} -- a formal proof of the Pythagorean-Triple 
      Theorem using the SAT-solver Glucose (2015).%
      \footnote{The Pythagorean-Triple Theorem asserts: \emph{It is not possible
      to divide the set of positive integers into two subsets $A$ and $B$
      such that neither $A$ nor $B$ contains a Pythagorean triple.}
      A triple $(a,b,c)$ of positive integers is Pythagorean if $a^2+b^2=c^2$. 
      More on this history at the 
      {\href{https://en.wikipedia.org/wiki/Boolean_Pythagorean_triples_problem}
      {\it Boolean Pythagorean triples problem}} webpage. A detailed presentation of 
      {\href{http://www.labri.fr/perso/lsimon/glucose/}{\it The Glucose SAT Solver}}
      can be found at its website.}
\end{itemize}
The five preceding formal proofs are not just proofs, but proofs that
come with a guarantee of correctness, the result of using logic-based
theorem-provers and interactive proof-assistants.  In the case of two
of these five theorems, the Four-Color Theorem and the Odd-Order Theorem,
there were in fact earlier proofs, but always suspected of containing
errors because of their length and complexity; these earlier proofs
had to be partly aided by \emph{ad hoc} computer programs, \ie, each
written for a specific purpose and themselves never verified to be
error-free.

In the case of the three other theorems mentioned above, they had
resisted all prior attempts, with or without the help of \emph{ad hoc}
computer programs. Their proofs were finally clinched only because of
advances in the underlying theory of theorem-provers and
proof-assistants (as well as, it must be stressed, improvements in the
speed and power of the hardware on which they were implemented).


These five formally-proved theorems by no means exhaust the list of 
theorems that have been formalized and mechanically proved with a 
correctness guarantee. I select them here because their
proofs resolved long-standing open problems in five
different areas of mathematics.%
  {
  \footnote{A list of $100$ theorems that have been proposed by researchers
  as benchmarks for theorem provers and proof assistants can be found on the Web at:
  {\href{http://www.cs.ru.nl/~freek/100/}{\em Formalizing 100 Theorems}}.
  Their adaptation to two advanced automated systems can be found at:
  {\href{https://madiot.fr/coq100/}{\em Formalizing 100 Theorems in Coq}},
  and {\href{http://www.cse.unsw.edu.au/~kleing/top100/}
  {\em The Top 100 Theorems in Isabelle}}. Of those that have
  been carried out so far, the vast majority are from the years after 2000.
  }}

I also single out for inclusion in my timeline 
(Figure~\ref{fig:third-twenty}) the emergence of 
the \emph{univalent foundations} of mathematics, largely
 spurred by the preceding development (of very large and
complicated proofs for simply-stated theorems which, if left to humans,
`will remain incomplete or contain errors with probability one'%
  \footnote{I am paraphrasing M.
  Aschbacher who wrote ``human beings are not capable of writing up a 10,000-page
  argument which is entirely free of errors. [$\ldots$] the 
  probability of an error in the proof is one''~\cite{aschbacher2005}. M.
  Aschbacher is a leading researcher in the
  classification of finite simple groups.}):
\begin{itemize}[itemsep=2pt,parsep=2pt,topsep=2pt,partopsep=0pt] 
\item \emph{Univalent foundations} of mathematics and
      \emph{homotopy type theory} (2006+).%
      \footnote{See the webpages {\href{https://homotopytypetheory.org/}
      {\em Homotopy Type Theory and Univalent Foundations}}
      and {\href{https://en.wikipedia.org/wiki/Homotopy_type_theory}
      {\em Homotopy Type Theory}} for more details.
      }
\end{itemize}
Although the early principles of \emph{univalent foundations} were
first formulated in the years 
from 2006 to 2009 with the specific goal of enabling the use of
automated proof-assistants to verify theorems and constructions in
classical mathematics, this new area has grown into a much larger body of
research in the foundations of mathematics -- and provides an
excellent illustration for how earlier logic-based developments in computer
science have subsequently triggered new and unforeseen directions in 
mathematics and mathematical logic.%
  \footnote{For an entertaining account, see Kevin Hartnett,
  ``Will Computers Redefine the Roots of Math?'' \emph{Quanta Magazine},
  19 May 2015 (available on the Web at 
  {\href{https://www.quantamagazine.org/20150519-will-computers-redefine-the-roots-of-math/}{\em Will Computers Redefine the Roots of Math?}}).}

I list several Turing Award winners in Figure~\ref{fig:third-twenty}
who were strongly influenced by logic and formal methods:
\begin{itemize}[itemsep=2pt,parsep=2pt,topsep=2pt,partopsep=0pt] 
\item Amir Pnueli (1996), for his work
   in \emph{temporal logic} and contributions to \emph{formal verification}.
\item E.M. Clarke, E.A. Emerson, and J. Sifakis (2007), for their work in 
   \emph{model-checking}.
\item Leslie Lamport (2013), for his work in distributed and concurrent systems.
\end{itemize}
Some may question my inclusion of Leslie Lamport in this
list. However, from my own reading, Lamport's work and innovations
(particularly the formal specification languages TLA and TLA+, the
basis of later implemented model checkers) were highly informed by
ideas about \emph{rewriting} and \emph{transition systems} and can
therefore be traced back to \emph{proof theory} (here in the form of
\emph{temporal logics}).%
  \footnote{It is also an assessment supported by the citation
  for Leslie Lamport at the
  {\href{http://amturing.acm.org/award_winners/lamport_1205376.cfm}
  {\it A.M. Turing Awards}} website.}

In Figure~\ref{fig:third-twenty}, under the column `Milestones/Accolades',
I also list:
\begin{itemize}[itemsep=2pt,parsep=2pt,topsep=2pt,partopsep=0pt] 
\item Alloy (1997), a model checker that has proved particularly successful
  in producing counter-examples.%
  \footnote{ {\href{http://alloy.mit.edu/alloy/}{\emph{Alloy: 
   a language \& tool for
   relational models}}}. From its webpage, ``the Alloy language is a simple
   but expressive logic based on the notion of relations, and was inspired by 
   the Z specification language and Tarski's relational calculus.''
   Like many other model checkers, Alloy is implemented on top of a SAT solver,
   \ie, Alloy works by reduction to a SAT solver, and is as good
   as the SAT solver it uses.} 
\item EasyCrypt (2009), a tool combining automated
  formal reasoning about relational properties of probabilistic
  computations with adversarial code, which has been successfully used 
  to verify game-based cryptographic proofs.%
      \footnote{ Details from its website, {\href{https://www.easycrypt.info/trac/}
                 {\it EasyCrypt: Computer-Aided Cryptographic Proofs}}.}
\item The CompCert project (2006), which produced a formally-certified compiler
      for the C programming language, using the proof-assistant Coq.%
      \footnote{An overview of the project is by its leader 
      Xavier Leroy~\cite{leroy2006,leroy2009}. Full details and updates
      are from the project website {\href{http://compcert.inria.fr/}{\it CompCert}}.
      More on Coq from its website {\href{https://coq.inria.fr/about-coq}
      {\it The Coq Proof Assistant}}.}
\item The seL4 project (2009), which verified an operating system micro-kernel
      with the automated proof-assistants Isabelle and HOL.%
      \footnote{An overview of the project is by G. Klein 
      \emph{et al}~\cite{klein2009}. More on the proof-assistants
      Isabelle and HOL from their respective websites,
      {\href{http://isabelle.in.tum.de/}{\it Isabelle}} and
      {\href{https://hol-theorem-prover.org/}{\it HOL}}.}
\item Certification of the FSCQ file system (2015), which uses proof-assistant
      Coq and logic-of-program Crash Hoare Logic 
      (an extension of Hoare Logic with a `crash' condition).%
      \footnote{More on FSCQ at {\href{http://css.csail.mit.edu/fscq/}{\it
       A Formally Certified Crash-Proof File System}} and the references
       therein.}
\end{itemize}
As the selection of these last five items reflects my own perspective,
they most certainly exclude other recent developments equally worthy of
mention, but which I know only by name. I have
used Alloy and compared it with other available automated tools in
graduate courses, and I have covered parts of CompCert and seL4 in 
another graduate course. It is worth noting that the
FSCQ project is only one of several which started in the last decade
or so and whose focus is on producing formally verified systems
software; all of them use Coq, Isabelle, or HOL, as automated
proof-assistant, together sometimes with an appropriate adaptation or 
extension of Hoare Logic.%
  \footnote{Some of these other projects are reviewed by the
    principals of the FSCQ project in their joint paper~\cite{chen2015}.}

 \Hide{   : Haogang Chen, Daniel Ziegler, Tej
    Chajed, Adam Chlipala, M. Frans Kaashoek, and Nickolai Zeldovich,
    ``Using Crash Hoare Logic for Certifying the FSCQ File System,''
    in \emph{Proceedings of Symp. on Operating Systems Principles}, Oct. 4-7,
    2015, Monterey, California, USA.}


\section{Timeline}
  \label{sect:timeline}

  My proposed timeline is in three parts, in
  Figures~\ref{fig:first-twenty}, \ref{fig:second-twenty},
  and~\ref{fig:third-twenty}.  

  I include events that say something significant about the
  interaction between the two fields, as well as events that are
  unrelated to this interaction in order to place the former in a
  wider context.  To distinguish the two kinds of events, I highlight
  those that are logic-related with a gray background.  The wider
  context helps understand the changing character of the interaction,
  as many parts of computer science become more formalized over the
  years, mediated by more levels above the hardware (actual physical
  computers, circuits, ethernets, etc.), and more focused on producing
  higher-level abstractions and software artifacts.

  I try not to `double list' events, \eg, not to list both the year of
  a discovery \emph{and} the year of its presentation in a
  professional journal or conference, and not to list both the year of
  an article \emph{and} the year (many years later) when that
  article's author receives public recognition. In all these cases, I
  choose to list the later year, not the earlier. This is, for
  example, the case of all the Turing Awards that are in my
  timeline.

  Of course, there are several other awards in computer science
  besides the Turing Awards, and which are named to honor the greats
  of mathematical logic. These include the Alonzo Church Award, the
  Kleene Award, and the G\"{o}del Prize.%
  \footnote{More information on these awards from their respective websites:
  the \href{http://siglog.org/awards/alonzo-church-award/}{\it Alonzo Church Award},
  the \href{http://lics.siglog.org/archive/kleene-award.html}{\it Kleene Award},
  and the \href{https://www.sigact.org/Prizes/Godel/}{\it G\"{o}del Prize}.
  }
  To keep the timeline within bounds, however, I limit attention to Turing
  Awards and, further, among the latter I only select those bearing an
  explicit direct influence from mathematical logic (as I see it) -- 
  and these are only a small sample of the pervasive influences
  between the two fields. 

\newpage


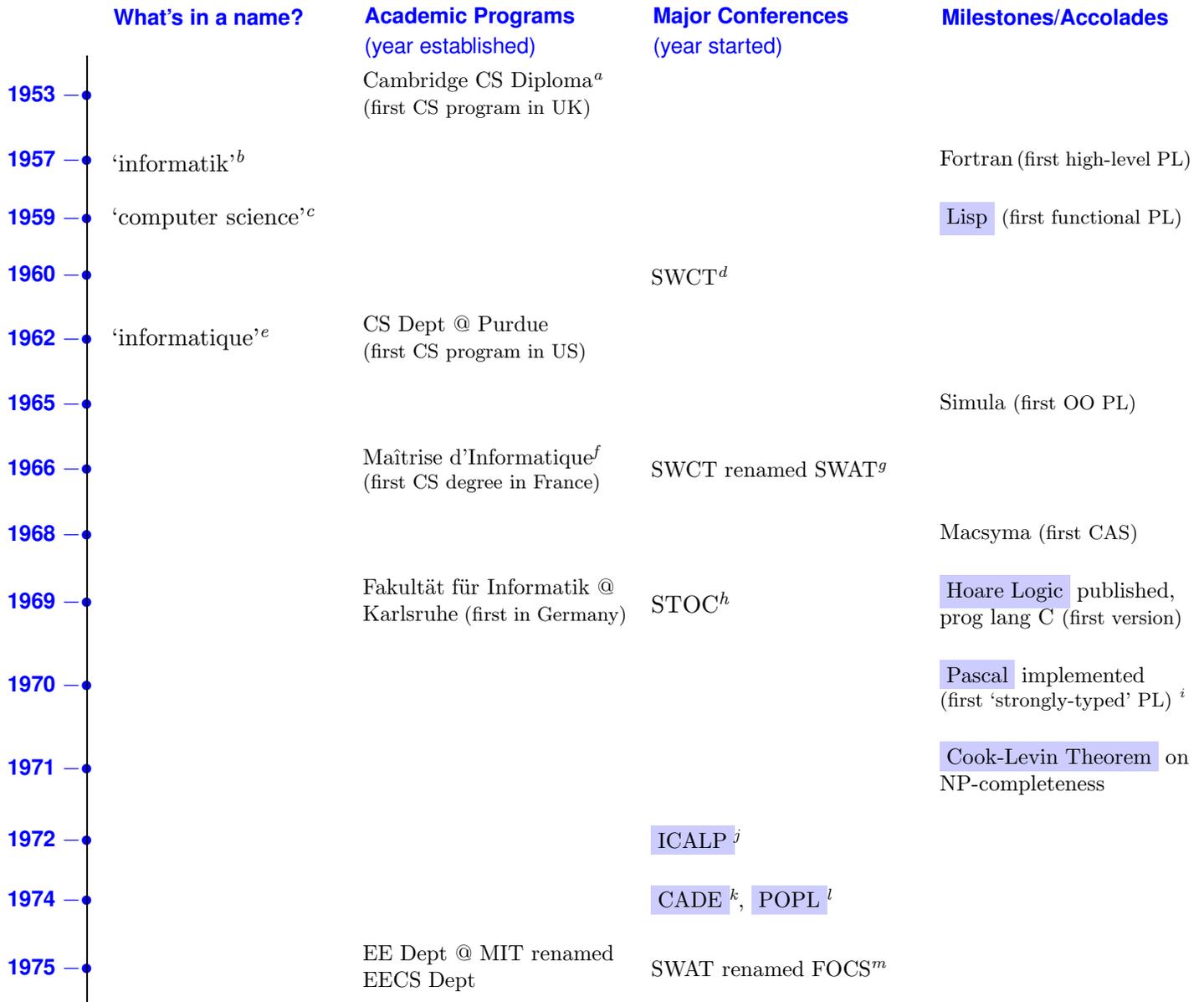
{ 
\begin{figure}[H]
\begin{custommargins}{-1.0cm}{-2cm} 
\hspace{-10pt}
\begin{minipage}{1.12\textwidth}
\noindent
\Headings{\textbf{What's in a name?}\\ }
   {\textbf{Academic Programs}\\ (year established)}
   {\textbf{Major Conferences}\\ (year started)} 
   {\textbf{Milestones/Accolades}\\ }
\\[-2ex]   
\yrTL{1953}{}
 {\small Cambridge CS Diploma\footnote{Official name: \emph{Diploma
  in Numerical Analysis and Automatic Computing}, claimed
  ``world's first full-year taught course in CS.''} 
 {\footnotesize (first CS program in UK)}}{}{}
\yrTL{1957}{`informatik'\footnote{First use of the term
    in German, by Karl Steinbuch~\cite{steinbuch1957}.}}
    {}{}{\small Fortran\footnotesize{\,(first high-level PL)}}
\yrTL{1959}{`computer science'\footnote{First use of the term
    in a CACM article, by Louis Fein~\cite{fein1959}.}}{}{}
    {\small \HItxt{Lisp} \footnotesize{(first functional PL)}}
\yrTL{1960}{}{}{\small SWCT\footnote{Annual Symposium on 
    \emph{Switching Circuit Theory and Logical Design} (SWCT), first held
    in Chicago, Illinois.}}{}
\yrTL{1962}{`informatique'\footnote{First use of the term in French, by
    Philippe Dreyfus~\cite{dreyfus1962}. 
    Other terms besides `computer science' and `informatics' were
    proposed: Some survived (`computing science', 
    `datalogy' in Scandinavia), others disappeared (`comptology',
    `hypology', `computology')  .}}
    {\small CS Dept @ Purdue \\
    {\footnotesize (first CS program in US)}}{}{}
\yrTL{1965}{}{}{}{\small Simula \footnotesize{(first OO PL)}}
\yrTL{1966}{}{\small Ma\^{i}trise d'Informatique\footnote{Promoted by the
    French government's \emph{Plan Calcul}. More on this at a
    \href{https://fr.wikipedia.org/wiki/Plan_Calcul}{Wikipedia entry} 
    (in French).}\\ \footnotesize{(first CS degree in France)}}
    {\small SWCT renamed SWAT\footnote{Annual Symposium on 
    \emph{Switching and Automata Theory}, first held (as SWAT)
    in Berkeley, California.}}{} 
\yrTL{1968}{}{}{}{\small Macsyma \footnotesize{(first CAS)}}
\yrTL{1969}{}
 {\small Fakult\"{a}t f\"{u}r Informatik @ Karlsruhe
 {\footnotesize (first in Germany)}}
  {STOC\footnote{\emph{Symposium on the Theory of Computing} (STOC), first
   held in Marina del Rey, California.}}
  {\small \HItxt{Hoare Logic} published,\\ 
  prog lang C {\footnotesize(first version)}}
\yrTL{1970}{}{}{}{\small \HItxt{Pascal} implemented
         \\ \footnotesize{(first `strongly-typed' PL)
            \footnote{Pascal is `almost' but not quite strongly-typed, and
            the first PL allowing higher-order (limited to second order
            in actual implementations)
            structured datatypes and procedures.
            See the critique by B.W. Kernighan~\cite{kernighan1981} who tends
            to minimize Pascal's innovations. 
            }}}
\yrTL{1971}{}{}{}{\small \HItxt{Cook-Levin Theorem} on NP-completeness}
\yrTL{1972}{}{}{\small \HItxt{ICALP}\footnote{\emph{Int'l Colloquium on Automata,
    Languages, and Programming} (ICALP), first held in Paris, France.
    ICALP was held in 1972 first, in 1974 a second time, and annually since 1976. 
    ICALP is highlighted, along with CADE and POPL,
    because of its Track B coverage.}}{}
\yrTL{1974}{}{}{\small \HItxt{CADE}\footnote{Mostly Biennial
    \emph{Conference on Automated Deduction}, 
    first held in Argonne National Lab, Illinois.},
    \HItxt{POPL}\footnote{Annual Symposium on
    \emph{Principles of Programming Languages} (POPL), 
    first held in Boston, Massachusetts.}
    }{}
\yrTL{1975}{}{\small EE Dept @ MIT renamed EECS Dept}
    {\small SWAT renamed FOCS\footnote{Annual Symposium on
    \emph{Foundations of Computer Science}, first held (as FOCS)
     in Berkeley, California.}}{}
\vspace{-0.11in}
\end{minipage}
\caption{The First Two Decades 
    (significant logic-related moments highlighted).} 
\label{fig:first-twenty}
\end{custommargins}
\end{figure}
}

\clearpage 

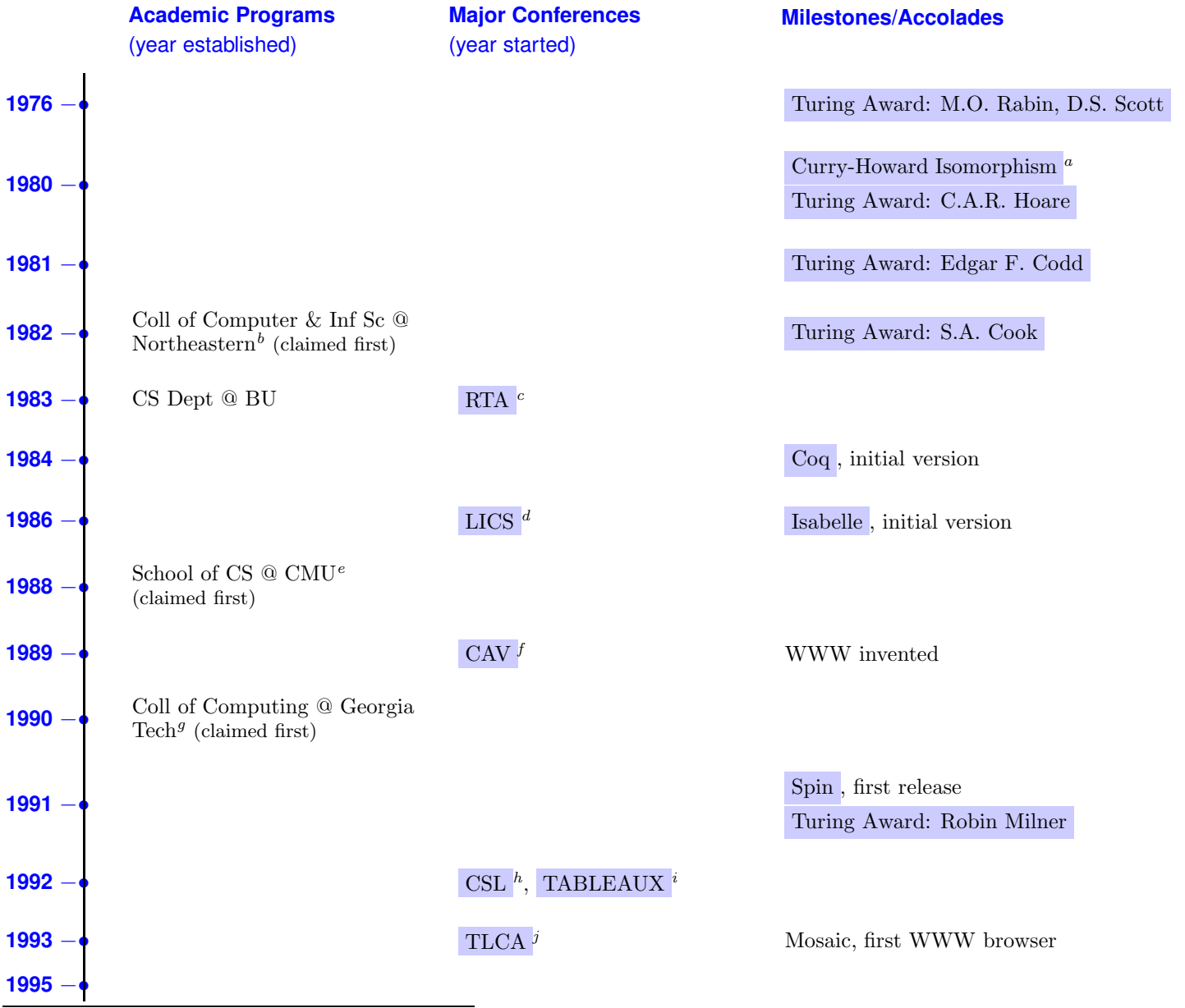
\begin{afterpage}
{
\begin{figure}[H]
\begin{custommargins}{-1.0cm}{-2cm} 
\hspace{-16pt}
   \begin{minipage}{1.12\textwidth}
\noindent
\Headingss{\textbf{Academic Programs}\\ (year established)}
   {\textbf{Major Conferences}\\ (year started)}
   {\textbf{Milestones/Accolades}\\ }
\yrTLL{1976}{}{}{\small\HItxt{Turing Award: M.O. Rabin, D.S. Scott}}
\yrTLL{1980}{}{}
 {\small \HItxt{Curry-Howard Isomorphism}\footnote{Curry published
   the Curry-Howard Isomorphism in 1958 in his~\cite{curry+feys1958},
   Section 9E, pp. 312-314. Howard's manuscript was written in 1969,
   published in 1980~\cite{howard1980}.}\\
   \HItxt{Turing Award: C.A.R. Hoare} }
\yrTLL{1981}{}{}{\small\HItxt{Turing Award: Edgar F. Codd}}   
\yrTLL{1982}{\small Coll of Computer \& Inf Sc @ Northeastern\footnote{More 
    at {\href{http://www.ccis.northeastern.edu/about/history/}{\emph{College of
    Computer and Information Science: Our History}}}.}
     \footnotesize{(claimed first)} 
     }{}
    {\small\HItxt{Turing Award: S.A. Cook}}
\yrTLL{1983}{\small CS Dept @ BU}{\small \HItxt{RTA}\footnote{Annual Conference on 
    \emph{Rewriting Techniques and Applications}, first held in Dijon, France.
    More at {\href{http://rewriting.loria.fr/rta/}{\it RTA Home Page}}.}}{}
\yrTLL{1984}{}{}
 {\small \HItxt{Coq}, initial version }
\yrTLL{1986}{}{\small \HItxt{LICS}\footnote{Annual Symposium on 
    \emph{Logic in Computer Science}, first held in Boston, Massachusetts.}}
    {\small \HItxt{Isabelle}, initial version }
\yrTLL{1988}{\small School of CS @ CMU\footnote{More 
  at {\href{https://www.csd.cs.cmu.edu/content/mission-history}
  {\emph{Computer Science at Carnegie Mellon: Mission and History}}}. 
  According to {\href{https://www.cs.cmu.edu/scs25/history}
  {\emph{CMU School of CS: 25th Anniversary}}}, it
  was ``the first college devoted solely to computer science in the 
  United States, and a model for others that followed.''}
  \\   \footnotesize{(claimed first)} 
     }{}{}
\yrTLL{1989}{}{\small \HItxt{CAV}\footnote{Annual Conference on 
    \emph{Computer Aided Verification}, first held in Grenoble, France.
    More at {\href{http://cavconference.org/}{\em CAV homepage}}.} }
    {\small WWW invented}
\yrTLL{1990}{\small Coll of Computing @ Georgia Tech\footnote{
   More at {\href{http://www.cc.gatech.edu/about/history}{\emph{History of
    GT Computing}}}.}
   \footnotesize{(claimed first)} 
   }{}{}
\yrTLL{1991}{}{}
 { \small \HItxt{Spin}, first release\\
 \HItxt{Turing Award: Robin Milner} }
\yrTLL{1992}{}{\small  \HItxt{CSL}\footnote{\href{http://www.eacsl.org/}
    {\emph{Computer Science Logic} annual conferences}, organized by the European
     Association for CSL. }, 
    \HItxt{TABLEAUX}\footnote{Int'l Conf 
    on Automated Reasoning with Analytic \emph{Tableaux} and Related Methods, 
    first held in Karlsruhe, Germany.}}{}
\yrTLL{1993}{}
  {\small \HItxt{TLCA}\footnote{Biennial Annual Conference on 
  \emph{Typed Lambda Calculi and Applications.}}}
  {\small Mosaic, first WWW browser}
\yrTLL{1995}{}{}{}
\vspace{-0.18in}
  \end{minipage}
\caption{The Second Two Decades 
    (significant logic-related moments highlighted).} 
\label{fig:second-twenty}
\end{custommargins}
\end{figure}
}
\end{afterpage}

\clearpage 

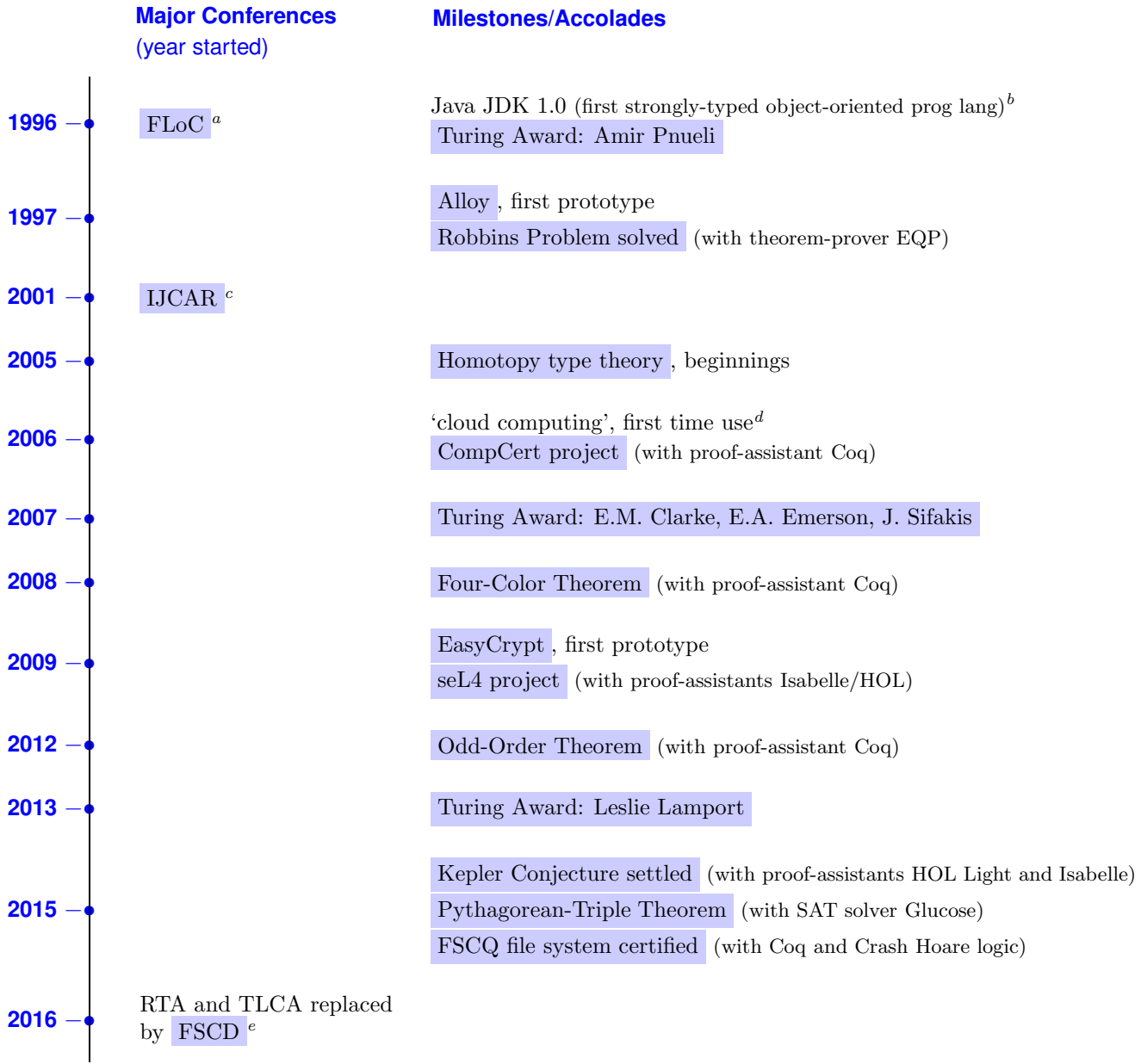
\begin{afterpage}
{
\begin{figure}[H]
\begin{custommargins}{-1.0cm}{-2cm} 
\hspace{-10pt}
  \begin{minipage}{1.12\textwidth}
\noindent
\Headingsss
   {\textbf{Major Conferences}\\ (year started)} {\textbf{Milestones/Accolades}\\ }
\yrTLLL{1996}{\small \HItxt{FLoC}\footnote{The \emph{Federated Logic Conference} 
 (FLoC) is held roughly every four years. FLoC is now the premier 
 international conglomeration of several mathematical logic and 
 computer science conferences that deal with the intersection of the 
 two fields.}}{\small Java JDK 1.0 
    {\footnotesize{(first strongly-typed object-oriented prog lang)}}\footnote{Java 
               JDK 1.0 was developed by J. Gosling, M. Sheridan, and P. Naughton, 
               since 1991, released in January 1996.} \\
               \HItxt{Turing Award: Amir Pnueli}}
\yrTLLL{1997}{}{\small \HItxt{Alloy}, first prototype \\
        \HItxt{Robbins Problem solved} \footnotesize{(with theorem-prover EQP)} }
\yrTLLL{2001}{\small \HItxt{IJCAR}\footnote{The 
 \emph{Int'l Joint Conference on Automated Reasoning} (IJCAR) is
 held semi-regularly every two-to-four years. IJCAR is a
 conglomeration of several conferences: CADE, FTP, TABLEAUX, and others.
 More information at {\href{http://www.ijcar.org/}{\emph{Home -- IJCAR}}}.}}{}
\yrTLLL{2005}{}{\small \HItxt{Homotopy type theory}, beginnings}
\yrTLLL{2006}{}{\small `cloud computing', first time use%
    \footnote{There is a debate about
    who was the first to coin the expression and when.
    However, in the current sense of a paradigm in which users 
    increasingly access software and computer power over the Web instead 
    of on their desktops, it seems the expression was first used on August 9,
    2006, when then Google CEO Eric Schmidt introduced it to an industry 
    conference. } \\
    \HItxt{CompCert project} \footnotesize{(with proof-assistant Coq)} }
\yrTLLL{2007}{}{\small \HItxt{Turing Award: E.M. Clarke, E.A. Emerson, J. Sifakis}}
\yrTLLL{2008}{}{\small
     \HItxt{Four-Color Theorem} \footnotesize{(with proof-assistant Coq)}}
\yrTLLL{2009}{}{\small
     \HItxt{EasyCrypt}, first prototype \\
     \HItxt{seL4 project} \footnotesize{(with proof-assistants Isabelle/HOL)} }
\yrTLLL{2012}{}{\small
     \HItxt{Odd-Order Theorem} \footnotesize{(with proof-assistant Coq)}}
\yrTLLL{2013}{}{\small \HItxt{Turing Award: Leslie Lamport}}
\yrTLLL{2015}{}{\small
     \HItxt{Kepler Conjecture settled} 
           {\footnotesize{(with proof-assistants HOL Light and Isabelle)}}\\
     \HItxt{Pythagorean-Triple Theorem} {\footnotesize{(with SAT solver Glucose)}}\\
     \HItxt{FSCQ file system certified} 
     {\footnotesize{(with Coq and Crash Hoare logic)}}
     }
\yrTLLL{2016}{\small RTA and TLCA replaced\\ 
   by \HItxt{FSCD}\footnote{ {\href{http://fscdconference.org/}
   {Int'l Conference on \emph{Formal Structures} for \emph{Computation}
    and \emph{Deduction}}}.}}{}
  \end{minipage}
\caption{The Third Two Decades 
    (significant logic-related moments highlighted).} 
\label{fig:third-twenty}
\vspace{-0.05in}
\end{custommargins}
\end{figure}
}
\end{afterpage}

\clearpage


{
\clearpage 
\clearpage
}


\section{Concluding Remarks}
\label{sect:conclusion}

Stressing only the positive in past sections, I may have presented a
permanent picture of harmony and close collaboration between
mathematical logic and computer science. As a matter of fact, it has
not been always so. From time to time, a few blemishes have marred
this picture -- or maybe they are just a reflection
of a far-flung fast-growing field.

\paragraph{When computer scientists do not know what logicians
did already.}

I mention five examples in no particular order, a sample from the
earlier years of computer science, more than three decades ago
(purposely). There are doubtless many, old and new, of which I am
ignorant.
\begin{enumerate}[itemsep=2pt,parsep=2pt,topsep=2pt,partopsep=0pt] 
\item The \emph{polymorphic lambda calculus}, also known as \emph{System F}:  
   It was first formulated (and many of its deep properties
   established in relation to second-order logic) by the 
   logician Jean-Yves Girard in the years 1970-72. It was
   independently re-formulated, with different syntactic conventions,
   by the computer scientist John Reynolds and published in 1974.%
   \footnote{Girard's formulation and results appeared in print
   in~\cite{girard1972}, Reynolds' formulation appeared in~\cite{reynolds1974}.
   In a later article in which he expanded on his results
   about System F from the early 1970's, Girard observed that 
   ``the proofs of these results have been often redone in the current
   [computer science] literature''~\cite{girard1986}. }
\item The \emph{functions definable in the polymorphic lambda-calculus
   are exactly the recursive functions provably total in second-order
   arithmetic}: This result was first proved by 
   Jean-Yves Girard and published in 1971. It was proved
   again independently by Richard Statman and reported in 1981, spurred by
   other computer scientists' earlier inconclusive attempts.%
      \footnote{Pawel Urzyczyn pointed me to this discrepancy, which had been
      also noted by others, for example, by G\'{e}rard Huet in his lecture
      notes~\cite{huet1986} (end of Section 10.3.3).
      Girard's paper is~\cite{girard1971}, Statman's paper is~\cite{statman1981},
      a lucid presentation of the result
      is by S{\o}rensen and Urzyczyn~\cite{sorensen+urzyczyn2006} (Chapter 12).
      }
\item The \emph{pebble game}, an ubiquitous concept in many
   parts of computer science, which has undergone many variations and
   extensions over the years: The original simplest version is usually
   credited to Michael Paterson and Carl Hewitt, who defined it in
   1970, unaware of the logician Harvey Friedman's
   earlier formulation of the same idea (in a long and highly
   technical article). Friedman completed and published his article in
   1969.%
   \footnote{Though he did not call it the `pebble game',
   Friedman's formulation was in a report for the Logic
   Colloquium, held in Manchester in August 1969, and included in its
   proceedings~\cite{friedman1970}. I ran into Friedman's formulation
   by chance, while preparing for an article where I used the pebble
   game in several proofs~\cite{kfoury1983}.  Paterson's and Hewitt's
   original report was dated November 1970~\cite{paterson+hewitt1970}.
   Later, I wrote a follow-up article~\cite{kfoury1986} where I made
   explicit the correspondence between the Friedman version and the
   Paterson-Hewitt version, showing that the two were also used for
   the same purpose (informally, at the simplest level: programs accessing $n+1$
   storage locations can do computations that programs accessing only $n$
   storage locations cannot do, unless pairing functions are available).
   There are still today competing claims about the
   origins of \emph{pebbling} and the \emph{pebble game} which ought
   to be sorted out; see, for example, the websites
   {\href{https://en.wikipedia.org/wiki/Pebble_game}{\it Pebble Game}} and
   {\href{https://en.wikipedia.org/wiki/Graph_pebbling}{\it Graph
   Pebbling}}, neither of which, incidentally, mention Friedman's
   earlier work or Paterson's and Hewitt's. Part of the complication,
   it seems, is that \emph{pebbling} and the \emph{pebble game} and
   derived concepts are now used in separate areas of computer science,
   each with its own motivations and research community.  }
\item The \emph{algorithm to decide typability of terms in the simply-typed
   lambda-calculus}, often called the Hindley-Milner or
   Damas-Hindley-Milner algorithm:  A version 
   was defined and proved correct by Roger Hindley in the late 1960's,
   a related version was independently defined by Robin Milner in the late
   1970's, and the latter was re-written and proved correct by Luis
   Damas in 1984. This history was upended by Hindley in 2005
   when he discovered that Max Newman had been the first to develop an algorithm
   for the problem, and to prove it correct, in the early 1940's.%
   \footnote{The historical facts are recounted 
   in an article by Roger Hindley~\cite{hindley2008}, including all the pertinent
   references (his own paper, Milner's paper, Damas' paper, and
   Newman's paper). I am indebted to Pawel Urzyczyn for alerting me to
   Hindley's revised history of the typability algorithm. A
   complementary article, with additional discussion of
   Newman's version of the algorithm and its history, is by H. Geuvers and
   R. Krebbers~\cite{geuvers+krebbers2011}. }
   %
\item What is known as \emph{Newman's Lemma}, a fundamental result
   widely used by computer scientists dealing with combinatory
   reduction systems, including the lambda calculus: It states that
   `local confluence' of a notion of reduction, say
   $\vartriangleright$, implies `confluence' of $\vartriangleright$ if
   all $\vartriangleright$-reduction sequences are finite. Max Newman
   proved the lemma in 1942. 
   Though stated a little differently, the lemma was in essence anticipated and
   proved by Axel Thue some three decades earlier!%
   \footnote{I am indebted to Roger Hindley who directed me to the history of
   Newman's Lemma, reported in his history of the lambda
   calculus, co-authored with Felice Cardone~\cite{cardone+hindley2009};
   see in particular Section 5.2 on page 738 in that chapter, which includes
   all the pertinent references (Newman's article of 1942, Axel Thue's
   article of 1910, and an article by M. Steinby and W. Thomas from 2000
   that summarizes Thue's paper in English).}
\Hide{
\item \underline{From Pawel}:
   Regarding Engelfriet I meant his paper of Stoc 83.
   I have only recently realized that
   higher-order push-down stores are not due to Engelfriet but to
   a certain A.N. Maslov, a different person than the famous
   S.Yu.Maslov. (Just an example.)
   }
\end{enumerate} 
%
But these examples are just innocent misattributions or delayed
attributions, causing no more damage than some duplication of
effort. Far more serious is the situation with computer algebra systems
which were developed, since their beginnings in the mid-1960's,
outside logic and formal-methods concerns.
\Hide{
   \footnote{Yes, this also happens in other mathematical sciences,
   and even in the same area of the same discipline, but is it with
   the same frequency?}
   }

\paragraph{The situation with \emph{computer algebra systems} (CAS's).}

CAS's are another category of automated general-purpose systems
developed by computer scientists. They provide integrated environments
for the manipulation of mathematical expressions in algebra (\eg,
various kinds of optimizations), number theory (\eg, numerical
computations and series operations), analysis (\eg, differentiation
and integration), and other deeper areas of mathematics -- all very useful in
applications.%
  \footnote{Perhaps the current most popular CAS's are Mathematica and Maple.
   A comprehensive 
   {\href{https://en.wikipedia.org/wiki/List_of_computer_algebra_systems}
   {\it List of CAS's}} is available on the Web.  A valuable
   historical overview of CAS's is by Joel Moses, the lead developer
   of the first CAS, Macsyma~\cite{moses2012}.}

In contrast to automated theorem provers and interactive proof
assistants -- to which I gave the lion's share of accolades for the
third period in my timeline (Figure~\ref{fig:third-twenty}) -- popular
commercially-available CAS's are not built on principles of formal logic.
They do not carry out calculations according to formalized proofs or
satisfying formally-specified guarantees of correctness.  With no
formal safeguards available to the user, they have sometimes produced
obscure errors, difficult to trace and difficult to rectify.%
  \footnote{A disturbing example involving Mathematica was reported in
  a recent article~\cite{duran2014}. The authors accidentally
  discovered an error by comparing Mathematica's calculations with
  those of Maple. To determine which of the two CAS's was at fault,
  they had to run both on multiple randomly generated inputs; they did
  not have at their disposal formally specified conditions under which
  the CAS's can be safely used and return outputs with correctness
  guarantees. More disturbing still than an error whose source could
  not be identified or located (Mathematica and Maple are not
  open-source) was the fact that an earlier release (Mathematica 7)
  did not show the error, while later releases (Mathematica 9 and 10)
  did. }
%
And yet, despite their ``notorious unsoundness,'' CAS's are
``in widespread use in mathematics, and it is not always so easy to
explain away the lack of concern about their unsoundness.''%
   \footnote{I am quoting from an
   article by Alan Bundy~\cite{bundy2011}, a prominent advocate for the
   use of automated theorem-provers.  }

Be that as it may, there is now an increased awareness for the need to
build CAS's on stronger formal foundations.%
   \footnote{An approximate but useful distinction between the two
   categories is that, while automated theorem provers and proof
   assistants are `super search engines' (of formal proofs, built from
   axioms and deduction rules), CAS's are `super calculators' (mostly
   of numbers, derived from equations and formulas).}
This effort goes beyond earlier work of augmenting pre-existing CAS's
with logic-based functionalities or, conversely, augmenting
pre-existing theorem provers and proof assistants with CAS
functionalities.%
   \footnote{This earlier work is exemplified by various add-ons
       and interfaces, to connect the two sides without
       fundamentally re-designing either. Examples:
       Analytica with Mathematica~\cite{clarke+zhao1992},
       Isabelle with Maple~\cite{ballarin+et-al1995},
       Isabelle with the computer algebra library Sumit~\cite{ballarin1998},
       Theorema with Mathematica~\cite{buchberger+et-al2016},
       PVS with Maple and Mathematica~\cite{adams+et-al1999,adams+et-al2001}.}
The new effort is more principled in that it aims to combine the two
sides -- proof search and interactive proof-assistance on the one
hand, algebraic domain-specific computation on the other -- in
an integrated bottom-up formal design.%
     \footnote{In some ways, this more recent effort is akin to the earlier
     development of SMT solvers as extensions of SAT solvers, which
     was also an integrated development based on logic and formal
     methods. What is different now is that it aims to extend or combine in
     a single design more features and functionalities of advanced
     systems (for proof search, interactive proof-assistance, and
     domain-specific algebraic computation).  I include the following
     projects as examples of the more recent effort: the
     {\href{https://leanprover.github.io/}{\it LEAN Theorem Prover}} and its
     publications (available from its website),
     the {\href{http://focalize.inria.fr/}{\it FoCaLiZe}} project
     and its publications (available from its website), as well as individual
     contributions by others (\eg, the work of M.T. Khan and W.
     Schreiner~\cite{khan2012,khan2013} and some of Sicun Gao's recent work with his
     collaborators~\cite{gao+kong+clarke2013,gao+kong+clarke2013b,
     gao+kong+clarke2014,gao+zufferey2016}).
     }
This activity is still limited to a few research groups, but it gives
an inkling of what may yet become a new big frontier in the interaction
between mathematical logic and computer science.

\paragraph{Do pure mathematicians agree or care?} 

The optimism expressed in earlier sections about a growing mutual
dependence between computer science and mathematical logic -- and
mathematics in general -- is not shared by everyone. Many view the
good effects going in one direction only: That mathematics and its
formalisms underlie (or should underlie) much of computer science is
taken for granted, but that computer science may have (or will have)
an equally important impact of a different kind on mathematics is
taken as a dubious claim.  In fact, there appears to be an inbred
indifference or even resistance among many pure mathematicians,
notably many in the core traditional areas, to anything involving
computers in their own work beyond routine pedestrian tasks (Google
search, email, typesetting with LaTeX).%
\footnote{
 That attitude was more entrenched prior to the great breakthroughs of
 automated theorem provers and interactive proof assistants 
 (Section~\ref{sect:third-twenty}), with good reasons perhaps, given
 the checkered history of CAS's and the emergence of what is
 called \emph{experimental mathematics} since the early 1990's, which
 owes its existence to computers and carries a bad name among the
 traditionalists. David Mumford, a prominent mathematician and Fields
 Medalist, started his career in algebraic geometry before converting
 to an applied area of computer science (vision) in the 1980's.
 Having known practitioners on both sides of the divide,
 Mumford could write from experience 
 that ``the pure mathematical community by and large still regards
 computers as invaders, despoilers of the sacred ground'' (quoted
 in~\cite{horgan1993}).  More than two decades later, that divide and
 the debates it provokes persist, though less sharply. Consider, for
 example, what eminent number-theorist and algebraist Michael Harris
 has to say on this divide~\cite{harris2015}.}

A particularly damning remark was once made by Alexandre Grothendieck,
an eminent Fields Medalist and algebraic geometer. He objected to the
``purported proof [of the Four-Color Theorem], whose validity is no
longer based on a firm belief that derives from the understanding of a
mathematical situation, but rather on the trust that one is willing to
put in a machine that lacks the capacity to understand.''%
     \footnote{
     This is a loose translation of the original French:
     ``une `d\'{e}monstration' qui ne se trouve plus fond\'{e}e dans l'intime
     conviction provenant de la compr\'{e}hension d'une situation math\'{e}matique,
     mais dans le cr\'{e}dit qu'on fait \`{a} une machine d\'{e}nu\'{e}e de
     la facult\'{e} de comprendre,'' taken from the footnote on page 137
     of Grothendieck's unpublished manuscript~\cite{grothendieck1986}.
     }
This view was echoed by Pierre Deligne, another Fields Medalist: ``I
don't believe in a proof done by a computer [$\ldots$].  In a way, I
am very egocentric. I believe in a proof if I understand it, if it's
clear.''%
  \footnote{Quoted in~\cite{horgan1993}. Deligne's statement, as well
  as Grothendieck's, precede the great advances of
  theorem provers and interactive proof assistants since the late 1990's.}
But these were reactions to \emph{ad hoc} computer programs, written
for specific exhaustive searches (enormous beyond human ability), not
to the more recent breakthroughs resulting from the use of automated
logic-based systems (surveyed in Section~\ref{sect:third-twenty}).
Nevertheless, the idea that the latter can become instruments of
mathematical progress is still a minority view, rejected by many
(most?) pure mathematicians.%
   \footnote{Consider, for example, what another pure mathematician 
   says dismissively about logical formalisms and,
   by implication, automated logic-based systems in mathematics:
   ``My mathematics colleagues almost never think about mathematical
    logic. [$\ldots$] They simply learn not to make certain moves that lead to
    trouble (as long as the referee doesn't complain, what,
    me worry?). [$\ldots$] So mathematicians work informally and have always
    done so; there is almost no trace of mathematical logic in most of the
    history of modern mathematics.''~\cite{edwards2012} }

But change is coming. Dissenting views have been expressed
by eminent members of the pure mathematical community itself.%
   \footnote{Some are expressed in Michael Harris' blog on the 
   {\href{https://wordpress.com/read/blogs/82268879/posts/585}{\it Univalent
   Foundations}} program. There is a clear separation between mathematicians
   (mostly against) and computer scientists (all in favor); some of
   the former are wavering and a few topologists even express strong
   support for the program.}
Most notable are Vladimir Voevodsky's, who made contributions to
core areas of pure mathematics (\eg, \emph{motivic homology}
and \emph{cohomology}, for which he received the Fields Medal) and,
since around 2005 and until his untimely death in 2017, to the foundations of
automated interactive proof-assistants (\emph{univalent foundations}
and \emph{homotopy type theory}).

{
\clearpage 
\clearpage
}


\renewcommand{\refname}{\sc references} 

\renewcommand{\bibpreamble}{{\small 
    \begin{quote}
      My list of references is not a bibliography. It is limited to
      references I used to justify my timeline.  I make no claim of
      fairness in my selection.  Except for a handful of historical
      character which I consulted for this article, and another
      handful supplied by colleagues who read earlier drafts, all
      the other citations are from books, articles, and webpages, that
      I have referenced or required in courses I have taught over
      nearly four decades. (And, no, I didn't read them all from
      cover to cover! In most of the books, I only read very, very few
      sections of particular interest to me.)
    \end{quote}
     }}

\renewcommand{\bibfont}{\footnotesize} 

{
\bibliographystyle{plain} 
\bibliography{./logic}
}

\end{document}